\documentclass[12pt]{article}
\usepackage[]{graphicx}
\usepackage[]{color}
\usepackage{natbib}
\usepackage{texments}
\usepackage{verbatim}
\usepackage[colorlinks=true,
linkcolor=red,
urlcolor=blue,
citecolor=cyan]{hyperref}
\usepackage[margin=1.2in]{geometry}

\makeatletter
\def\maxwidth{ %
	\ifdim\Gin@nat@width>\linewidth
	\linewidth
	\else
	\Gin@nat@width
	\fi
}
\makeatother

\newcommand{\tC}{\widetilde{\bC}}

\definecolor{fgcolor}{rgb}{0.345, 0.345, 0.345}

\usepackage{framed}
\makeatletter
{\par\unskip\endMakeFramed%
	\at@end@of@kframe}
\makeatother

\definecolor{shadecolor}{rgb}{.97, .97, .97}
\definecolor{messagecolor}{rgb}{0, 0, 0}
\definecolor{warningcolor}{rgb}{1, 0, 1}
\definecolor{errorcolor}{rgb}{1, 0, 0}

\usepackage{alltt}

\usepackage{thumbpdf,lmodern}

\usepackage{framed}

\newcommand{\blue}[1]{\leavevmode\textcolor{black}{#1}}
\usepackage{amsmath}
\usepackage{amssymb}
\usepackage{amsfonts}
\usepackage{subfigure}
\usepackage{booktabs}
\usepackage{caption}
\usepackage{multirow}

\newcommand{\bA}{\textbf{A}}
\newcommand{\bb}{\textbf{b}}
\newcommand{\bB}{\textbf{B}}

\newcommand{\bC}{\textbf{C}}

\newcommand{\bD}{\textbf{D}}

\newcommand{\bF}{\textbf{F}}

\newcommand{\bh}{\textbf{h}}

\newcommand{\bI}{\textbf{I}}

\newcommand{\bL}{\textbf{L}}

\newcommand{\bP}{\textbf{P}}
\newcommand{\bs}{\textbf{s}}

\newcommand{\bu}{\textbf{u}}

\newcommand{\bv}{\textbf{v}}

\newcommand{\bw}{\textbf{w}}

\newcommand{\bx}{\textbf{x}}
\newcommand{\bX}{\textbf{X}}
\newcommand{\by}{\textbf{y}}
\newcommand{\bY}{\textbf{Y}}
\newcommand{\br}{\textbf{r}}

\newcommand{\bZ}{\textbf{Z}}
\newcommand{\bz}{\textbf{z}}

\newcommand{\mbs}[1]{\boldsymbol{#1}}

\newcommand{\bSigma}{\mbs{\Sigma}}

\newcommand{\bpsi}{\mbs{\psi}}

\newcommand{\bbeta}{\mbs{\beta}}

\newcommand{\blambda}{\mbs{\lambda}}

\newcommand{\bgamma}{{\mbs{\gamma}}}
\newcommand{\btheta}{{\mbs{\theta}}}

\renewcommand{\det}[1]{\text{det}\left(#1\right)}

\newcommand{\ben}{\begin{equation*}}
\newcommand{\een}{\end{equation*}}
\newcommand{\bean}{\begin{eqnarray*}}
	\newcommand{\eean}{\end{eqnarray*}}
\newcommand{\bsm}{\begin{smallmatrix}}
	\newcommand{\esm}{\end{smallmatrix}}
\newcommand{\bmat}{\begin{matrix}}
	\newcommand{\emat}{\end{matrix}}

\newcommand{\bzero}{\textbf{0}}

\newcommand{\given}{\,|\,}
\newcommand{\taus}{\tau^2}
\newcommand{\sigs}{\sigma^2}
\newcommand{\eps}{\epsilon}

\newcommand{\tSig}{\widetilde \bSigma}
\newcommand{\boeta}{ \mbox{\boldmath $ \eta $} }

\newcommand{\iid}{\overset{iid}{\sim}}
\newcommand{\ind}{\overset{ind}{\sim}}

\title{\blue{Nearest-neighbor} sparse Cholesky matrices in spatial statistics}
\author{Abhirup Datta\footnote{Email: abhidatta@jhu.edu}\\
	\small{Department of Biostatistics, Johns Hopkins University}}
\date{}
\begin{document}
	\maketitle
	
	
\begin{abstract}
Gaussian Processes (GP) is a staple in the toolkit of a spatial statistician. Well-documented computing roadblocks in the analysis of large geospatial datasets using Gaussian Processes have now largely been mitigated via several recent statistical innovations. Nearest Neighbor Gaussian Processes (NNGP) has emerged as one of the leading candidate for such massive-scale geospatial analysis owing to their empirical success. This articles reviews the connection of NNGP to sparse Cholesky factors of the spatial precision (inverse-covariance) matrix. Focus of the review is on these sparse Cholesky matrices which are versatile and have recently found many diverse applications beyond the primary usage of NNGP for fast parameter estimation and prediction in the spatial (generalized) linear models. In particular, we discuss applications of sparse NNGP Cholesky matrices to address multifaceted computational issues in spatial bootstrapping, simulation of large-scale realizations of Gaussian random fields, and extensions to non-parametric mean function estimation of a Gaussian Process using Random Forests. 
We also review a sparse-Cholesky-based model for areal (geographically-aggregated) data that addresses long-established interpretability issues of existing areal models. Finally, we highlight some yet-to-be-addressed issues of such sparse Cholesky approximations that warrants further research.
\end{abstract}

	
	\section{Introduction} \label{sec:intro}
	Spatially-indexed data are commonly encountered in many diverse fields of research including forestry, climatology, environmental health, ecology, infectious disease epidemiology, neuro-imaging, etc. The objective of spatial statistics has primarily been to develop models and algorithms that can utilize the location information in the data to improve statistical inference. In particular, spatial (generalized) linear models using Gaussian Processes (GP) have become a staple tool for such geospatial analysis. 
	Let $\bs_1, \bs_2, \ldots, \bs_n$ denote $n$ locations and $y(\bs_i)$ and $\bx(\bs_i)$ respectively denote the univariate response and a $p\times 1$ vector of covariates at the $i^{th}$ location. A linear mean GP model for such geo-spatial data specifies $y(\cdot) \sim GP(\bx'(\cdot)\bbeta, \Sigma(\cdot,\cdot))$ where $\bx'(\cdot)\bbeta$ is the linear mean function and $\Sigma(\cdot,\cdot)$ is the covariance function \citep{banerjee2014hierarchical,cressie2015statistics}. This endows the responses with a linear mean $E(y(\bs))=\bx(\bs)'\bbeta$ and the covariance $Cov(y(\bs),y(\bs'))=\Sigma(\bs,\bs')$. If $\by=(y(\bs_1),\ldots,y(\bs_n))'$ denotes the vector of observed responses and $\bX$ denotes the corresponding design matrix created by stacking up the $\bx(\bs_i)$'s, then the GP model implies  
	\begin{equation}\label{eq:gpmodel}
	\by \sim N(\bX\bbeta,\bSigma) \mbox{ where }  \bSigma=(\Sigma(\bs_i,\bs_j)). 
	\end{equation}
	The resulting multivariate Gaussian likelihood
	\begin{equation}\label{eq:gplik}
	p(\by) \propto \frac 1{\sqrt{\det{\bSigma}}}\exp\Big(-\frac 12 (\by - \bX\bbeta)'\bSigma^{-1}(\by - \bX\bbeta)\Big),
	\end{equation}
	is used to infer about $\bbeta$ and the parameters in the spatial covariance function $\Sigma$. Computing this likelihood 
	involves storing the $n \times n$ covariance matrix $\bSigma$, requiring $O(n^2)$ storage. It also involves computing the inverse and the determinant of $\bSigma$. These operations are typically performed via the Cholesky decomposition and requires $O(n^3)$ computation time or floating point operations (FLOPs). Neither the storage or computing demands of a GP likelihood can be afforded by typical personal computers even for moderately large $n$ ($\sim 10^4$ or larger). 
	
	Over the last few decades the spatial statistics community has attacked this big GP problem from many fronts and offered many different and efficient solutions to ease the computational burden. Methods include sparse nearest neighbor approximations \citep{ve88,stein2004,nngp}, low-rank approximations \citep{banerjee2008gaussian,cressie2008fixed},  sparse-plus-low-rank method \citep{ma2019additive}, multi-resolutional approaches \citep{katzfuss2017multi,guhaniyogi2017large}, data partitioning \blue{and divide-and-conquer approaches \citep{barbian2017spatial,guhaniyogi2017divide,guhaniyogi2018meta,guhaniyogi2019multivariate,guhaniyogi2020distributed},} covariance tapering \citep{furrer2006covariance,kaufman2008covariance}, stochastic partial differential equations \citep{lindgren2011explicit}, composite likelihoods \citep{bevilacqua2015comparing,eidsvik2014estimation}, grid-based methods \citep{nychka2015multiresolution,guinness2017circulant,stroud2017bayesian}, among others. A comprehensive review of all these methods is beyond the scope of this paper but we refer the readers to the articles \cite{sun2012geostatistics,bradley2016comparison,banerjee2017high,heaton2019case,Banerjee2020} for reviews and comparisons of the methods. 
	
	In this manuscript, we focus on reviewing Nearest Neighbor Gaussian Processes (NNGP) \citep{nngp} based on Vecchia's approximation \citep{ve88}. This strand of literature, reviewed in Section \ref{sec:rev}, itself has become sufficiently large owing to the empirical success of this method. 
	The traditional use of this method has primarily been fast parameter estimation and prediction for geospatial data in a GP-based linear model. As discussed above there are many alternative methods to achieve these tasks and the recent data analysis comparisons of \cite{heaton2019case} demonstrated that many of the methods achieve highly competitive performance. However, the NNGP method naturally yields sparse Cholesky factors of the spatial precision (inverse-covariance) matrices which have a wider range of applications beyond parametric estimation and prediction in the spatial linear model.
	This review focuses primarily on some such novel applications of these sparse Cholesky factor matrices in spatial statistics. 
	
	In Section \ref{sec:rec}, we first expand on 3 recent applications of these nearest-neighbor sparse Cholesky matrices in geospatial data --- resampling or bootstrap for spatial data (Section \ref{sec:boot}), large-scale simulation of Gaussian random fields (Section \ref{sec:sim}), and non-parametric mean function estimation for Gaussian Processes using random forests (Section \ref{sec:rf}). We demonstrate how sparse Cholesky matrices are used to resolve multi-faceted computational issues in each of these applications. Our fourth and final example (Section \ref{sec:areal}) digresses from the setting of geospatial (point-referenced) data, and considers areal (geographically-aggregated) data. We review a new class of models using neighbor-based sparse Cholesky factors for such data that improves over existing approaches in terms of parameter interpretability and model performance while remaining computationally scalable. We conclude in Section \ref{sec:disc} with a discussion of two yet-to-be-addressed aspects of the NNGP method that open up future avenues of research. 
	
	
	
	
	
	
	\section{Nearest-neighbor based sparse modeling of large spatial data}\label{sec:rev}
	
	\subsection{Vecchia's nearest-neighbor approximation} 
	
	In a seminal paper, \cite{ve88} proposed leveraging the spatial structure encoded in GP covariance functions to obtain a computationally scalable approximation of the GP likelihood (\ref{eq:gplik}). The approach proceeds by rewriting (\ref{eq:gplik}) as 
	\begin{equation}\label{eq:telescope}
	\begin{aligned}
	p(\by) & = p(y(\bs_1))p(y(\bs_2) \given y(\bs_1))\ldots p(y(\bs_n) \given y(\bs_1), \ldots, y(\bs_{n-1})) \\
	& = p(y(\bs_1)) \prod_{i=2}^n p(y(\bs_i) \given \by(H(\bs_i))), 
	\end{aligned}
	\end{equation}
	where $H(\bs_i)=(\bs_1,\ldots,\bs_{i-1})'$ and $\by(H(\bs_i))$ is the vector formed by stacking $y(\bs)$ for $\bs \in H(\bs_i)$. Most popular choices of the covariance function $\Sigma$ conforms to the `first law of geography', i.e., ensures that proximal things are more similar than distant ones. The magnitude of the spatial covariance decays with distance and the conditioning sets $H(\bs_i)$ will contain many members far away from $\bs_i$ and thus offering minimal information about it. \cite{ve88} proposed approximating (\ref{eq:telescope}) with 
	\begin{equation}\label{eq:nntele}
	p(y(\bs_1)) \prod_{i=2}^n p(y(\bs_i) \given \by(N(\bs_i)))
	\end{equation}
	where $N(\bs_i)$ is a set of up to $m$ nearest neighbors of $\bs_i$ chosen from $H(\bs_i)$, and $\by(N(\bs_i))$ is defined similar to $y(H(\bs_i))$. For two sets $A,B \subseteq \{1,\ldots,n\}$, let $\bSigma(A,B)$ denote the sub-matrix of $\bSigma$ with rows and columns respectively indexed by $A$ and $B$. Then under the GP model (\ref{eq:gpmodel}), the nearest neighbor likelihood (\ref{eq:nntele}) reduces to 
	\begin{small}
		\begin{equation*}
		\begin{aligned}
		N(y(\bs_1) \given \bx(\bs_1)'\bbeta, \bSigma(\bs_1,\bs_1))  \prod_{i=2}^n N(y(\bs_i) \given \bx(\bs_i)'\bbeta + \bb_i'(\by(N(\bs_i)) - \bX(N(\bs_i))\bbeta), f_i),
		\end{aligned}
		\end{equation*}
	\end{small}
	where $\bb_i =   \bSigma(N(\bs_i),N(\bs_i))^{-1}\bSigma(N(\bs_i),\bs_i)$, 
	$f_i = \bSigma(\bs_i,\bs_i) - \bSigma(\bs_i,N(\bs_i))\bb_i $, and $\bX(N(\bs_i))$ is the design matrix corresponding to $y(N(\bs_i))$.
	Parameter estimation proceeds by maximizing the pseudo-likelihood (\ref{eq:nntele}). The restriction of the {\em neighbor sets} $N(\bs_i)$ to be comprised of at most $m$ locations ensures we only need to store and invert $m \times m$ matrices $\bSigma(N(\bs_i),N(\bs_i))$, thereby reducing the storage and computing requirements to evaluate the likelihood (\ref{eq:nntele}) respectively to $O(nm^2)$ and $O(nm^3)$ respectively. 
	
	\cite{stein2004} demonstrated that as (\ref{eq:nntele}) is the product of correctly specified conditional densities $p(y(\bs_i) \given y(N(\bs_i))$, the score function from (\ref{eq:nntele}) yields a set of unbiased estimating equations for the parameters. 
	They also generalized this approximation in several ways including using block conditional densities instead of the univariate ones, exploring choices of neighbor sets beyond nearest neighbors, and formulating a Restricted Maximum Likelihood (REML) approach for parameter estimation. 
	
	\subsection{Nearest neighbor Gaussian Processes} 
	
	The GP model (\ref{eq:gpmodel}) is conceived from the additive spatial regression model $E(y(\bs_i)) = \bx(\bs_i)'\bbeta + w(\bs_i)$ where $\bx(\bs_i)'\bbeta$ is the linear covariate effect and $w(\bs_i)$ is a smooth spatial effect explaining residual structured variation in $y(\bs_i)$. Modeling $w(\cdot)$ as a Gaussian process with zero-mean and covariance function $C$ and assuming additive Gaussian iid errors $\eps_i$ leads to (\ref{eq:gpmodel}) with the covariance function $\Sigma(\cdot,\cdot) = C(\cdot,\cdot) + \taus \delta(\cdot,\cdot)$ where $\taus$ is the error variance and $\delta$ is the white noise covariance function, i.e., $\delta(\bs,\bs')=I(\bs=\bs')$. Thus the GP regression (\ref{eq:gpmodel}) is the marginal form of the hierarchical mixed effect model
	\begin{equation}\label{eq:gphier}
	\begin{aligned}
	y(\bs_i) & = \bx(\bs_i)'\bbeta + w(\bs_i) + \eps(\bs_i), \\
	\bw & =(w(\bs_1),\ldots,w(\bs_n))' \sim N(0,\bC) \mbox{ where } \bC=(C(\bs_i,\bs_j)),\\
	\eps(\bs_i) &\iid N(0,\taus).  
	\end{aligned}
	\end{equation}
	
	The original nearest neighbor approximation of Vecchia was directly applied to the data likelihood (\ref{eq:gplik}) for $\by$. Often, inference on the latent spatial surface $w(\bs)$ is of interest for scientists to understand structured variation in the response beyond what is explained by the covariates. 
	\cite{nngp} generalized the idea of \cite{ve88} from a data likelihood approximation to {\em `Nearest Neighbor Gaussian Processes'} (NNGP) --- a new valid class of multivariate Gaussian distributions and Gaussian random fields that can be used to conduct fast spatial inference on observed or latent processes.  
	The key to this extension is the following connection of Vecchia's nearest neighbor approximation to sparse Cholesky matrices. 
	
	For any Gaussian Process $w(\cdot)$ with zero mean and covariance function $C$, akin to (\ref{eq:nntele}), the nearest neighbor approximation of the likelihood of  realizations of the process $w(\cdot)$ at $\bs_1,\ldots,\bs_n$ is given by
	\begin{equation}\label{eq:nntelew}
	\begin{aligned}
	p(w(\bs_1)) \prod_{i=2}^n p(w(\bs_i) \given \bw(N(\bs_i)))\;.
	\end{aligned}
	\end{equation}
	The term $p(w(\bs_i) \given \bw(N(\bs_i)))$ is the conditional density 
	$N(w(\bs_i) \given \bb_i'\bw(N(\bs_i)), f_i)$ 
	where 
	\begin{equation}\label{eq:bfw}
	\begin{aligned}
	\bb_i = \bC(N(\bs_i),N(\bs_i))^{-1}\bC(N(\bs_i),\bs_i)\\
	f_i = \bC(\bs_i,\bs_i) - \bC(\bs_i,N(\bs_i))\bb_i
	\end{aligned}
	\end{equation}
	and can be considered as the likelihood from the generative model $w(\bs_i) = \bb_i'\bw(N(\bs_i)) + N(0,f_i)$. Thus the expression in (\ref{eq:nntelew}) can be equivalently written as the likelihood from the model:
	\begin{equation}\label{eq:nngpgen}
	\begin{aligned}
	w(\bs_1)=&\,\eta_1 \\
	w(\bs_2)=&\, b_{21}w(\bs_1) + \eta_2\\
	w(\bs_3)=&\, b_{31}w(\bs_1) + b_{32}w(\bs_2) + \eta_3\\
	\ldots \qquad & \ldots \\
	w(\bs_n)=&\, b_{n1}w(\bs_1) + b_{n2}w(\bs_2) + \ldots + b_{n,n-1}w(\bs_{n-1}) + \eta_n, \\
	\end{aligned}
	\end{equation}
	where $\eta_i \ind N(0,f_i)$ with $f_1=C(\bs_1,\bs_1)$, and $b_{ij}=0$ if $\bs_j$ is not a neighbor of $\bs_i$, and is the $l^{th}$ element of $\bb_i$ if $\bs_j$ is the $l^{th}$ neighbor of $\bs_i$. 
	We can stack the equations in (\ref{eq:nngpgen}) to have the matrix equation:
	\begin{equation}\label{eq:nngpgenmat}
	\bw = \bB \bw + \boeta
	\end{equation}
	where $\bw=(w(\bs_1),\ldots,w(\bs_n))'$, $\boeta=(\eta_1,\ldots,\eta_n)' \sim N(\bzero,\bF)$ with $\bF=\mbox{diag}(f_1,\ldots,f_n)$, and $\bB=(b_{ij})$ is a strictly lower triangular matrix. From (\ref{eq:nngpgenmat}), we have 
	\begin{equation}\label{eq:chol}
	(\bI - \bB)\bw = \boeta \iff \bw = (\bI - \bB)^{-1}\boeta \sim N(0,(\bI - \bB)^{-1}\bF(\bI - \bB)^{-T}).
	\end{equation}
	Thus \cite{nngp} noted that the nearest neighbor approximation of \cite{ve88} corresponds to a generative multivariate Gaussian model $\bw \sim N(0,\tC)$ for the process realizations, where $\tC=(\bI - \bB)^{-1}\bF(\bI - \bB)^{-T}$. This new model essentially replaces the model $\bw \sim N(\bzero, \bC)$ where $\bC=C(\bs_i,\bs_j \given \btheta)$, which corresponds to the full GP likelihood. 	
	The precision matrix 
	\begin{equation}\label{eq:nngpprec}
	\tC^{-1}=(\bI - \bB)'\bF^{-1}(\bI - \bB)
	\end{equation}
	 admits a Cholesky decomposition $\bL'\bL$ with the lower-triangular Cholesky factor \begin{equation}\label{eq:cholmat}
	\bL=\bF^{-1/2}(\bI-\bB).
	\end{equation} 
	As $\bB$ has at most $m$-non-zero elements per row and $\bF^{-1/2}$ is a diagonal matrix, the \blue{lower-triangular} Cholesky factor $\bL=(l_{ij})$ also has at most $m$ \blue{(sub-diagnoal)} non-zero elements per row. Computing $\bL$  involves only computing the $\bb_i$'s and $f_i$'s from (\ref{eq:bfw}) thus requiring $O(nm^3)$ FLOPs and $O(nm^2)$ storage as opposed to $O(n^3)$ FLOPs and $O(n^2)$ storage for computing the Cholesky factor of the full GP matrix $\bC$. Subsequent to computing $\bL$, the likelihood for $\bw$ involves computing quadratic forms and determinant of  $\tC^{-1}$. This is straightforward as any quadratic form $\bu'\tC^{-1}\bv=(\bL\bu)'(\bL\bv)$ and due to row-sparsity of $\bL$, the multiplication $\bL\bx$ only uses $O(nm)$ additional FLOPs. Similarly, $\det\tC=\prod_i l_{ii}^{-2}$. Therefore, computing the whole likelihood only requires linear (in $n$) storage and time.
	
	The generative approach of \cite{nngp} using sparse Cholesky factor has several benefits beyond the data likelihood approximation of \cite{ve88}. One can consider any hierarchical spatial model but simply replace the Gaussian Process prior for spatial random effects $\bw$ with an NNGP prior.
	For example, in (\ref{eq:gphier}), as $y(\bs_i)$'s are independent conditional of $\bw$ and $\bbeta$, 
	the joint likelihood 
	\begin{equation}\label{eq:hierlik}
	N(\by \given \bX\bbeta + \bw, \taus \bI) \times N(\bw \given \bzero, \tC)
	\end{equation} 
	can be evaluated efficiently even for large spatial data. 
	Augmenting this with priors for the other parameters ($\bbeta$, $\taus$, and parameters of the covariance function $C$), facilitates standard Bayesian inference on the latent effects $\bw$. One can also proceed with frequentist estimation using the EM algorithm by treating $\bw$ as the missing data. 
	
	Another benefit of NNGP is prediction of the latent process at new locations. \cite{nngp} specified the conditional distribution of $w(\bs_0) \given \bw$ at new locations $\bs_0 \notin S=\{\bs_1,\ldots,\bs_n\}$ is given by 
	\begin{equation}\label{eq:nnpred}
	\begin{aligned}
	w(\bs_0) \given \bw \ind &N(\bC(\bs_0,N(s_0))\bC(N(\bs_0),N(\bs_0))^{-1}\bw,\\ &C(\bs_0,\bs_0) - \bC(\bs_0,N(\bs_0))\bC(N(\bs_0),N(\bs_0))^{-1}\bC(N(\bs_0),\bs_0))
	\end{aligned}
	\end{equation}
	
	The prediction distribution is basically equivalent to kriging independently (conditional on $\bw$) at each new location $\bs_0 \notin S$ using $m$-nearest neighbors $N(\bs_0)$ of $\bs_0$ in $S$, instead of all of $S$.
	\cite{vecchia1992new} considered similar nearest neighbor-based kriging but only for point predictions as opposed to entire prediction distributions. \cite{katzfuss2020vecchia} extended the independent nearest-neighbor kriging to joint kriging which improved prediction quality. 
	
	\cite{nngp} demonstrated that Equations (\ref{eq:chol}) and (\ref{eq:nnpred}) complete the specification of a valid Gaussian Process over the entire domain which was referred to as the Nearest Neighbor Gaussian Process (NNGP). In any hierarchical model, NNGP can be used to replace GP prior to ensure fast computation by leveraging the sparse Cholesky factor $\bL$, and proceed with fast and full Bayesian inference on all parameters and the latent process $w(\cdot)$, and Bayesian predictive inference on the outcome process $y(\cdot)$. More detailed reviews of the method are available in \cite{wiresnngp} and \cite{banerjee2017high}. 
	
	\subsection{Recent work}\label{sec:litrev}
	There has been considerable recent work related to Vecchia's approximation and NNGP. \cite{gramacy2015local} developed a `{\em Local approximation GP}' which extends nearest-neighbor based kriging equations to non-stationary covariance functions. More classes of non-stationary covariance models with nearest neighbor approximations have been implemented recently in \cite{Risser2019}.  \cite{stroud2017bayesian} used the NNGP precision matrix $\tC^{-1}$ as a pre-conditioner for the full GP matrix $\bC$  to solve for unknowns $\bu$ in linear equations of the form $\bC \bu = \bz$ arising in conditional GP simulations. \cite{schafer2020sparse} demonstrated that, given the ordering and the neighbor sets, the Cholesky factor $\bL$ is the optimal one in terms of Kullback-Leibler distance between the full GP distribution and a sparse Cholesky based distribution. 
	
	\blue{Given an ordering, we can construct a NNGP model by directly specifying the Cholesky factor. Hence, regardless of what ordering we begin with, the sparsity of the Cholesky factor $\bL$ is exactly controlled by only considering $m$ directed nearest neigbhors under that ordering. However,} the quality of the nearest neighbor approximation and the resulting sparse Cholesky factor depends on the choice of data ordering. As spatial data does not have any natural ordering, simple orderings like sorting along some co-ordinate is often adopted. \blue{Empirical results have shown robustness of NNGP prediction performance to different co-ordinate based orderings \citep{nngp}.}  \cite{guinness2018permutation} studied alternate choices of ordering and observed that certain well-principled orderings or sometimes even random orderings can lead to more efficient parameter estimation.  \cite{katzfuss2021} considered joint orderings of the responses $\by$ and the latent random effects $\bw$ in a large class of models which they referred to as the `{\em Sparse Generalized Vecchia (SGV)}'. They demonstrated that the original Vecchia approximation for the response process and NNGP for the latent process (and many other popular spatial models) can be unified under the umbrella of SGV as they arise from imposing different relative orderings of $\by$ and $\bw$. 
	
	There has been substantial investigation on relative merits of using this approximation on the response process \citep{ve88} and the latent process \citep{nngp}. Endowing the latent process $\bw$ with an NNGP prior and 
	marginalizing $\bw$ out, yields:
	\begin{equation}\label{eq:resplik}
	\by \sim N(\bX\bbeta, \tC + \taus \bI).
	\end{equation}
	While the hierarchical model (\ref{eq:hierlik}) is equivalent to the marginal model (\ref{eq:resplik}), unfortunately, the latter is not directly amenable to scalable computing. This is because even if $\tC^{-1}$ inverse has a sparse Cholesky factor, the same cannot be said of $(\tC+\taus \bI)^{-1}$. The Bayesian implementation of \cite{nngp} avoids this issue by sampling the latent $\bw$ sequentially from (\ref{eq:hierlik}) in a Gibbs sampler. However, for large data, sequential sampling of $n$ latent random effects substantially increases the MCMC dimension and can lead to convergence issues. Motivated by these sampling issues \blue{of the sequential NNGP algorithm, 
	\cite{finley2019efficient} explored various strategies. A {\em collapsed NNGP} was considered that uses the marginal NNGP model (\ref{eq:resplik}) and leverages matrix identities combined with sparse matrix operations to speed up computations involving $(\tC+\taus \bI)^{-1}$. They also considered a {\em response NNGP} model assigning NNGP prior to directly model the response process,} facilitating a Bayesian model-based analogue of Vecchia's original likelihood approximation. \blue{ \cite{finley2019efficient} also implemented {\em Conjugate NNGP}, an MCMC-free version of the response NNGP model which uses a hybrid of exact Bayesian inference (for parameters having conjugate priors) and cross-validation (for other parameters).} However, \cite{katzfuss2021} has showed that using NNGP on the latent process leads to a better approximation of the full GP than when using Vecchia's approximation on the response process.
	\cite{zhang2019practical} has proposed a solution that retains this advantage of using the latent NNGP while circumventing the sequential sampling of $\bw$. They resourcefully use the sparse Cholesky factor of $\tC^{-1}$ in a Bayesian conjugate descent approach that enjoys the aforementioned benefits of the latent NNGP model ((\ref{eq:hierlik}) or (\ref{eq:resplik})) while enabling fast block sampling of $\bw$.
	\blue{ \cite{schafer2020sparse} has also proposed a fast algorithm for using the model in (\ref{eq:resplik}) with the NNGP-like prior on the latent process, using two sparse Cholesky decompositions -- one of $\tC^{-1}$ and one of $\tC^{-1} + \tau^{-2} \bI$.}
	
	Multiple software implementing Vecchia's approximation and Nearest Neighbor Gaussian Processes are now publicly available including CRAN R-packages \texttt{spNNGP} \citep{spnngp}, \texttt{BRISC} \citep{briscpkg}, \texttt{GPvecchia} \citep{gpvecchia}, \texttt{GpGp} \citep{gpgp}, and \texttt{BayesNSGP} \citep{BayesNSGP}. Many extensions of the method have also been developed including generalizations to spatio-temporal settings \citep{jones1997models,datta2016nonseparable}, multivariate settings \citep{taylor2019spatial}, non-Gaussian outcomes \citep{finley2020r,zilber2019vecchia}, spatio-temporal filtering \citep{jurek2020hierarchical}, and multivariate cumulative distribution functions of Gaussian Processes \citep{nascimento2020vecchia}.  
	Most of these have focused on parametric modeling, estimation and prediction using the spatial GP linear model. In the next Section, we discuss some novel applications of the NNGP sparse Cholesky factors. 
	
	
	
	\section{New applications}\label{sec:rec}
	\subsection{Bootstrapping of spatial data}\label{sec:boot}
	This Section reviews a fast parametric bootstrap developed in \cite{brisc} for inference (interval estimates) on parameters in the spatial regression model (\ref{eq:gpmodel}) or (\ref{eq:gphier}). The parameters consist of the regression coefficient $\bbeta$, the noise variance $\taus$ and the parameters $\btheta$ specifying the covariance function $C$. To obtain interval estimates of the parameters, 
	one can opt for a Bayesian implementation, sampling using MCMC methods from the joint-likelihood (\ref{eq:hierlik}) augmented by priors for the parameters  \citep{banerjee2014hierarchical}. These provide full posterior distributions of all the parameters from which one can obtain point and interval estimates of any function of the parameters. This strategy was adopted in the latent- \citep{nngp,zhang2019practical} and response- \citep{finley2009} NNGP models. However, the sequential nature of MCMC typically require many thousand iterations leading to prolonged analysis times despite the speedup per likelihood evaluation afforded by use of NNGP. 
	
	One can also leverage asymptotic distribution of the parameter MLEs (maximum likelihood estimates) in a frequentist setup. For spatial data, there are two paradigms of asymptotics. For `{\em infill asymptotics}' where data are observed with increased density in a fixed spatial domain, individual parameters may not be consistently estimable and only certain functions of the spatial parameters are identifiable \citep{chen00,zhang2004inconsistent,tang2019identifiability}. The results for the `{\em increasing domain}' setting where the spatial domain expands along with increased sample sizes, are more in line with traditional asymptotics.  \cite{mardia1984maximum} established asymptotic normality of parameters for a wide choice of covariance functions. However, asymptotic interval estimates rely on the asymptotic covariance of the parameters and these involve the Hessian of the full GP likelihood which is computationally infeasible for large $n$. The issue persists even when using the nearest-neighbor approximation (\ref{eq:nntele}) to the likelihood as one needs to use a computationally onerous sandwich-variance estimator 
	\citep{stein2004}. 
	
	A general alternative method for finite sample inference on parameters is bootstrapping. Bootstrapping can be done in an embarrassingly parallel fashion thereby easing some of the computational burden of the aforementioned approaches. However, spatial data are correlated violating the fundamental principle of resampling iid units used in bootstrap. Even if one wishes to proceed with bootstrap, ignoring this correlation, it is unclear how to define the GP covariance matrix for a bootstrapped dataset as the correlation between two resamples of the same data unit $(y,\bx,\bs)$ will be one as they will have the same location. 
	
	
	\cite{olea2011generalized} proposed a parametric  bootstrap for the spatial regression model (\ref{eq:gpmodel}). As Cov$(\by) = \bSigma$, we have Cov$(\bSigma^{-1/2}\by) = \bI$ where $\bSigma^{-1/2}$ denotes the Cholesky factor of the precision matrix $\bSigma^{-1}$. Thus, the residual vector $\br=\bSigma^{-1/2}(\by - \bX\bbeta)$ are iid $N(0,1)$ distributed and one can resample from them to create bootstrapped residual vectors $\br^{(1)},\ldots,\br^{(B)}$ where $B$ is the bootstrap sample size. Subsequently, generating $\by^{(b)}=\bX\bbeta + \bSigma^{1/2}\br^{(b)}$ for $b=1,\ldots,B$, creates the bootstrapped datasets having the same distribution as the original data. If $\bpsi$ denotes the entire set of parameters, one can run MLE for each dataset $\by^{(b)}$ in parallel to obtain the sample of MLEs $\bpsi^{(1)},\ldots,\bpsi^{(B)}$ from which one can derive  interval estimates. As $\bbeta$ and $(\btheta,\taus)$ (which parametrizes $\bSigma$) are unknown, in practice, they are replaced by their MLE for this algorithm. 
	
	The parametric bootstrap of \cite{olea2011generalized} requires the Cholesky factors of both $\bSigma$ and $\bSigma^{-1}$ which use $O(n^3)$ operations. Hence, despite being amenable to embarrassingly parallel computing, the algorithm cannot be used for large spatial datasets. \cite{brisc} noted that the replacing full GP with NNGP achieves both in $O(n)$ time. To see this, similar to (\ref{eq:chol}), let $\tSig$ denote the NNGP covariance matrix approximating $\bSigma$, i.e., 
	$\tSig=(\bI - \bB_y)^{-1}\bF_y(\bI - \bB_y)^{-T}$
	where $\bB_y$ and $\bF_y$ are defined respectively similar to $\bB$ and $\bF$, but for the covariance function $\Sigma$ of the response process $y(\cdot)$ instead of the covariance function $C$ of the latent process $w(\cdot)$. Hence akin to (\ref{eq:cholmat}), the Cholesky factor is  obtained as 
	\begin{equation}\label{eq:cholmaty}
	\bL_y := \tSig^{-1/2}=\bF_y^{-1/2}(\bI-\bB_y)
	\end{equation} 
	using linear (in $n$) storage and time. 
	As $\tSig^{-1/2} \approx \bSigma^{-1/2}$, we have Cov$(\tSig^{-1/2}\by) \approx \bI$. So \cite{brisc} proposed using the latter to decorrelate $\by$. 
	
\begin{figure}[]
		\centering
		\subfigure[Sample covariance of raw data from exponential GP]{\includegraphics[scale=0.23,trim={0 0 0 30},clip]{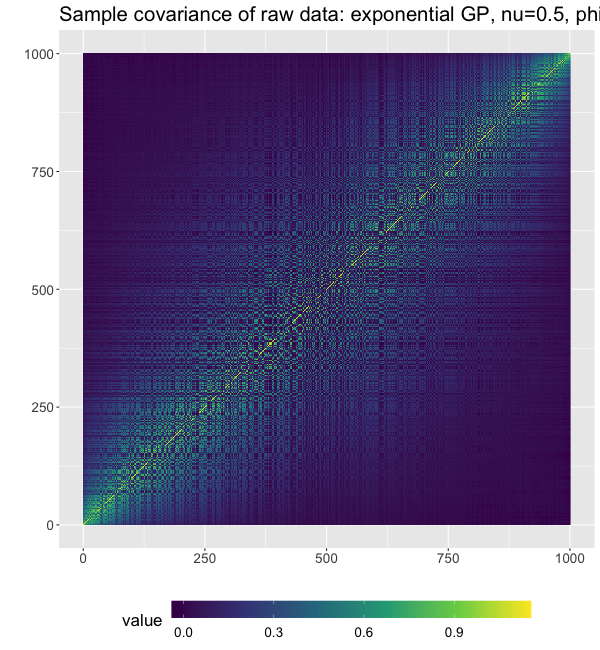}}
		\subfigure[Sample covariance NNGP-decorrelated data simulated from exponential GP ]{\includegraphics[scale=0.23,trim={0 0 0 30},clip]{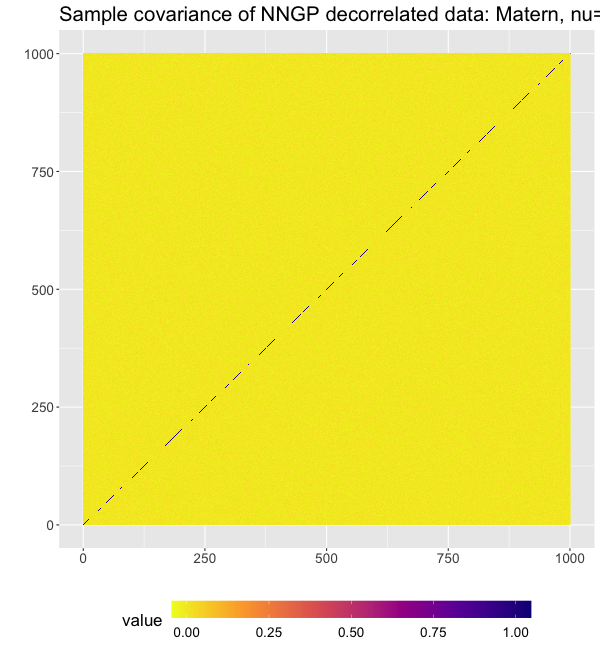}}
		\subfigure[Sample covariance NNGP-decorrelated data - $\bI$ ]{\includegraphics[scale=0.23,trim={0 0 0 30},clip]{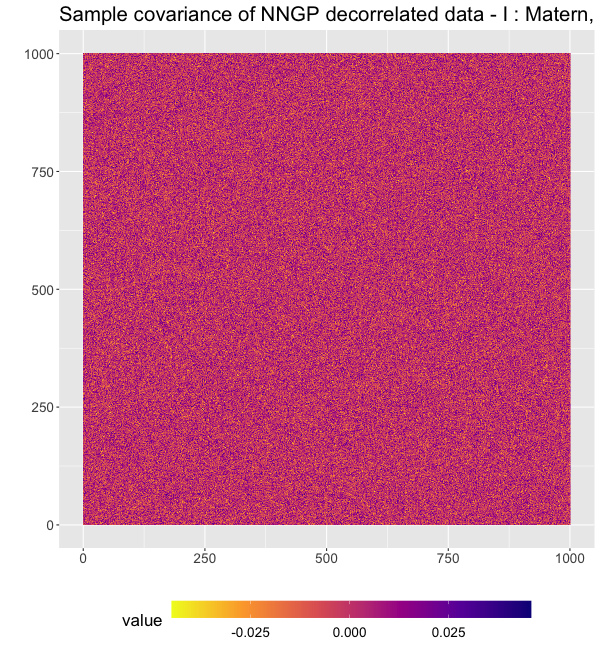}}
		\subfigure[Sample covariance of raw data from Mat\'ern-$3/2$ GP]{\includegraphics[scale=0.23,trim={0 0 0 30},clip]{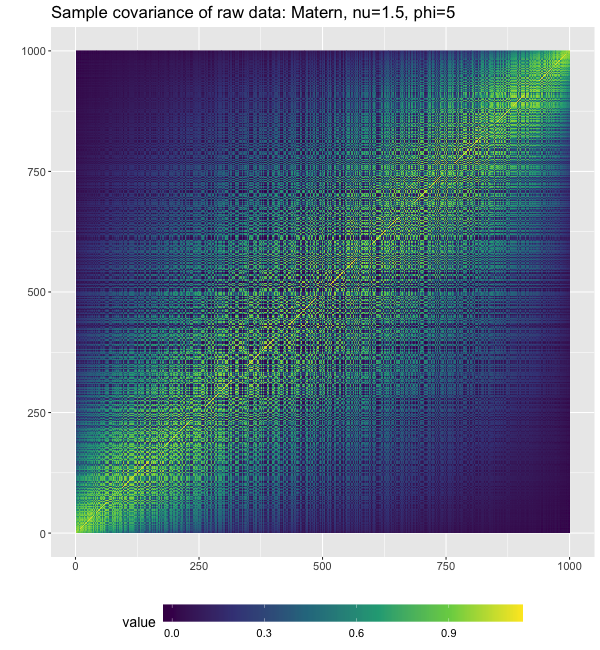}}
		\subfigure[Sample covariance NNGP-decorrelated data simulated from Mat\'ern-$3/2$ GP ]{\includegraphics[scale=0.23,trim={0 0 0 30},clip]{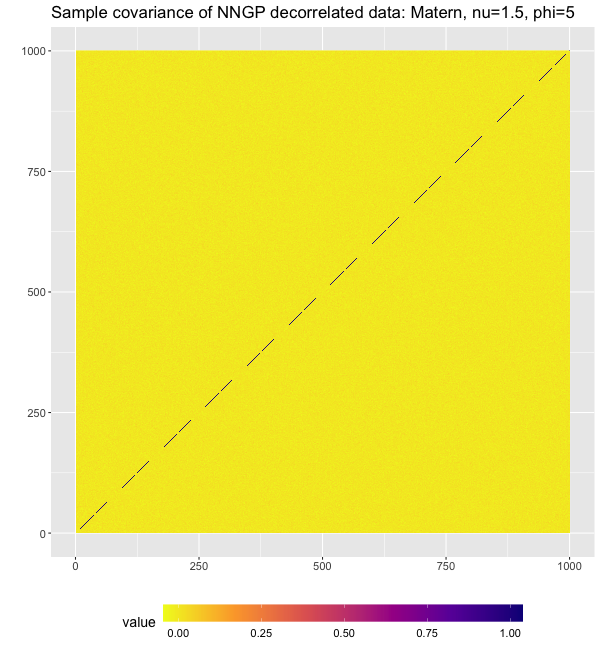}}
		\subfigure[Sample covariance NNGP-decorrelated data - $\bI$ ]{\includegraphics[scale=0.23,trim={0 0 0 30},clip]{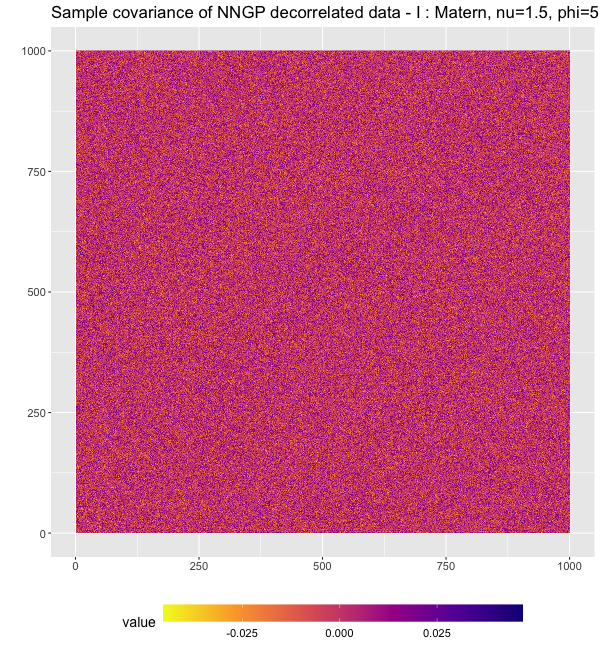}}
		\vskip-5mm\subfigure[Density of diagonal and off-diagonal entries of sample covariances of NNGP-decorrelated data]{\includegraphics[scale=0.5]{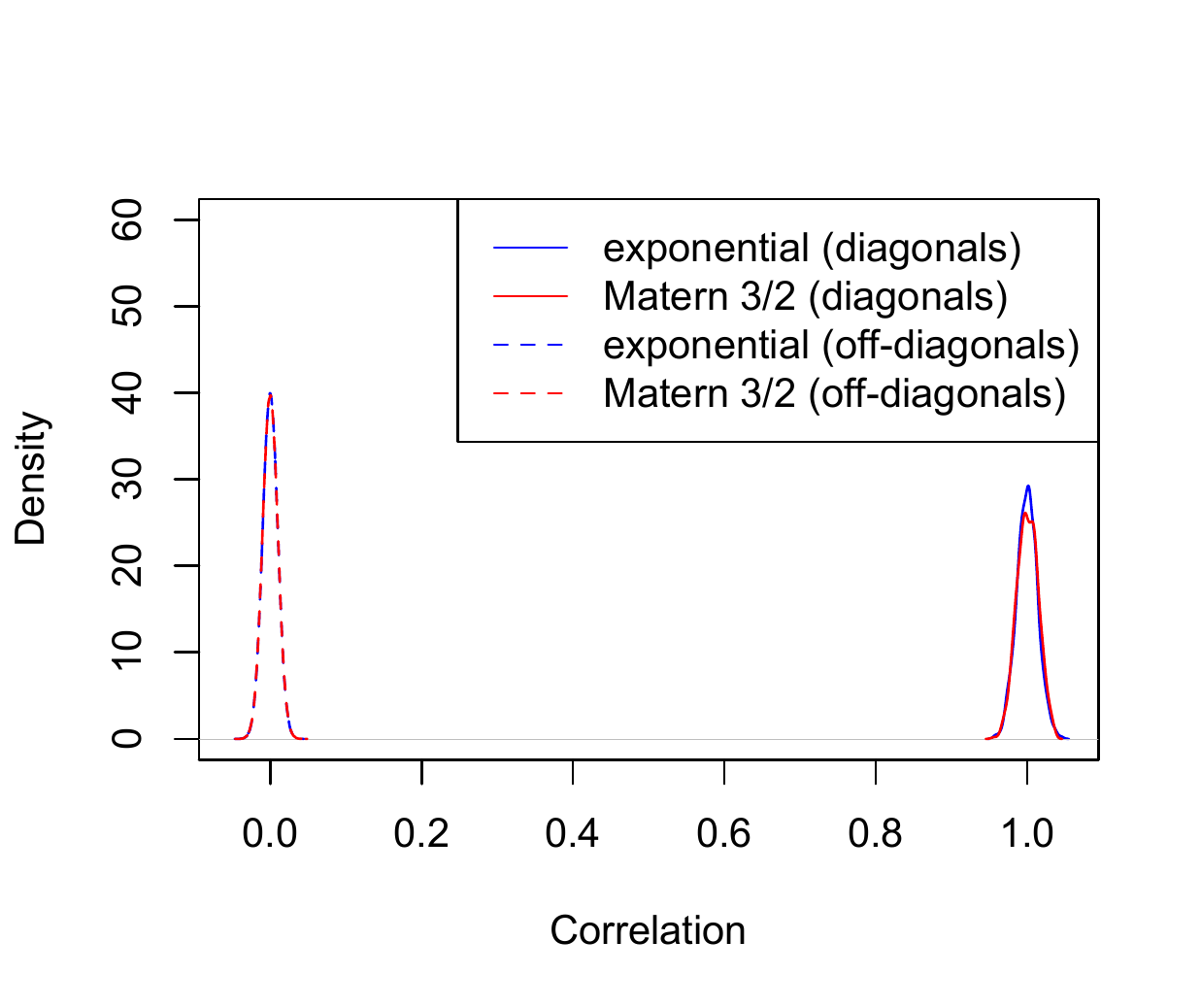}}
		\caption{Fast approximate decorrelation of GP-generated data using NNGP.} 
		\label{fig:decor}
	\end{figure}

	We provide a small illustration below to assess the quality of this  approximate decorrelation of spatial data generated from a full GP. We generated $1000$  locations randomly on a unit square and simulated $10000$ datasets each one being realization of a GP on those $1000$ locations. We used two choices of the covariance function: the exponential covariance function and the smoother Mat\'ern covariance with smoothness $3/2$ \citep{stein2012interpolation} which has a slower decay at small distances. For each choice, we first plotted the sample covariance matrix based on the $10000$ replicates in the left column. We then decorrelated the data using NNGP Cholesky factor $\tSig^{-1/2}$ from (\ref{eq:cholmaty}) and plotted the sample correlation matrix for the decorrelated vectors on the right column. We observe from Figure \ref{fig:decor} (middle and right) that for both choices of the covariance function, the sample covariance of the NNGP decorrelated datasets are close to the identity matrix. To further confirm this, in the bottom row we plot the densities of the diagonal and off-diagonal elements of the decorrelated covariance matrix. Once again for both choices of covariance functions, the density of the off-diagonal elements concentrates around $0$ while those for the diagonal elements are around $1$. Thus swapping the dense inverse Cholesky factor $\bSigma^{-1/2}$ with the NNGP analog $\tSig^{-1/2}$ achieves fast decorrelation without significant loss of information. 
	
	The second part of the bootstrap algorithm, requires correlating back the resampled errors $\br^{(b)}$ to get the bootstrapped datasets $\by^{(b)}=\bX\bbeta+\bSigma^{1/2}\br^{(b)}$. While it is natural to consider replacing $\bSigma^{1/2}$ with the NNGP approximation 
	\begin{equation}
	\tSig^{1/2} = \bL_y^{-1} = (\bI - \bB_y)^{-1}\bF_y^{1/2},
	\end{equation}
	unlike $\bL_y$, its inverse $\bL^{-1}_y$ is not directly available. Instead, \cite{brisc} propose an $O(n)$-time algorithm to compute products of the form $\bL_y^{-1}\bv$ required in the correlate-back step. Note that $\bu=\bL_y^{-1}\bv$ can be obtained by solving the triangular system $\bL_y\bu=\bv$ for $\bu$. Now $\bL_y=\tSig^{-1/2}$ has already been computed. From (\ref{eq:nngpgen}), we know that $\bL_y=(l^{(y)}_{ij})$ is lower-triangular with at most $m$ non-zero \blue{sub-diagonal} elements per row. Hence, one can back-solve for $\bu=(u_1,,\ldots,u_n)$ as follows:
	\begin{equation}\label{eq:backsolve}
	\begin{aligned}
	u_1&=v_1/l^{(y)}_{11} \\
	u_2&= (v_2 - l^{(y)}_{21}u_1)/l^{(y)}_{22} \\
	\ldots & \qquad \ldots \\
	u_n &= (v_n - \sum_{i < n :l^{(y)}_{ni} \neq 0} l^{(y)}_{ni}u_i)/l^{(y)}_{nn}.
	\end{aligned}
	\end{equation}
	The row-sparsity of $\bL_y$ ensures this only requires at most $O(nm)$ operations to obtain all the $u_i$'s. Thus the correlate-back step using NNGP is also linear in time and storage and does not require any matrix multiplication or inversion. \cite{brisc} referred to this fast bootstrap method as `{\em BRISC: Bootstrap for Rapid Inference on Spatial Covariances}' and empirically demonstrated how it provided well-calibrated interval estimates of all parameters in a GP regression while being manifold faster than MCMC based implementation of NNGP. The bootstrap method is implemented in the \texttt{BRISC\_bootstrap} 
	function of the \texttt{BRISC} CRAN R-package \citep{briscpkg}.
	
	\subsection{Simulating large spatial datasets using NNGP}\label{sec:sim}
	
	
	Simulating large datasets from a full GP is also \blue{generally}  computationally prohibitive. \blue{With the exception of compactly supported functions \citep{gneiting2002compactly,sanso1987finite} which yield sparse covariance matrices and facilitate efficient large-scale  simulation \citep{wood1994simulation,dietrich1997fast}, most common families of covariance functions like the exponential, Mat\'ern, or Gaussian are not compactly supported and for irregularly located spatial data, they yield a dense covariance matrix $\bSigma$ which does not possess any exploitable structure. Hence,} generating realizations from $N(\bzero,\bSigma)$ also involves either the Cholesky or the eigen-value decomposition of $\bSigma$. Both require $O(n^2)$ storage and $O(n^3)$ time, and cannot be accomplished using personal computing resources even for moderately large datasets. To illustrate, for a set of experiments detailed below, a personal computer with 16 GB of memory could not generate random draws from a full GP for $n \geq 10000$. 
	
	We here outline a simple and fast algorithm for simulation of large datasets from NNGP that provides an excellent approximation to draws from full GP. The idea is same as the correlate-back step of the bootstrap described in Section \ref{sec:boot}. As the NNGP covariance matrix $\tSig$ well approximates the full GP covariance $\bSigma$. If $\bz \sim N(0,\bI)$, we have $\by^{(sim)} = \tSig^{1/2}\bz \sim N(0,\tSig) \approx N(0,\bSigma)$. Thus, a random draw $\by^{(sim)}$ of size $n$ for an NNGP, only involves generating $n$ iid $N(0,1)$ draws $z_1,\ldots,z_n$ and computing the product $\tSig^{1/2}\bz$. As outlined in Section \ref{sec:boot}, this product can be obtained by first computing the reverse Cholesky factor $\bL_y=\tSig^{-1/2}$ and then solving for $\by^{(sim)}$ in the triangular system $\bL_y\by^{(sim)} = \bz$ using the sparse back-solve of (\ref{eq:backsolve}). \blue{The process requires $O(nm^3)$ FLOPs to calculate the Cholesky factor $\bL_y$ and $O(nm)$ to solve the sparse triangular system $\bL_y\by^{(sim)} = \bz$. The total storage required is $O(nm^2)$. Thus storage and time required for simulating realizations of NNGP remain linear in the sample size $n$.}
	
\begin{figure}
		\centering
		\subfigure[GP: Exponential]{\includegraphics[scale=0.23,trim={0 0 0 30},clip]{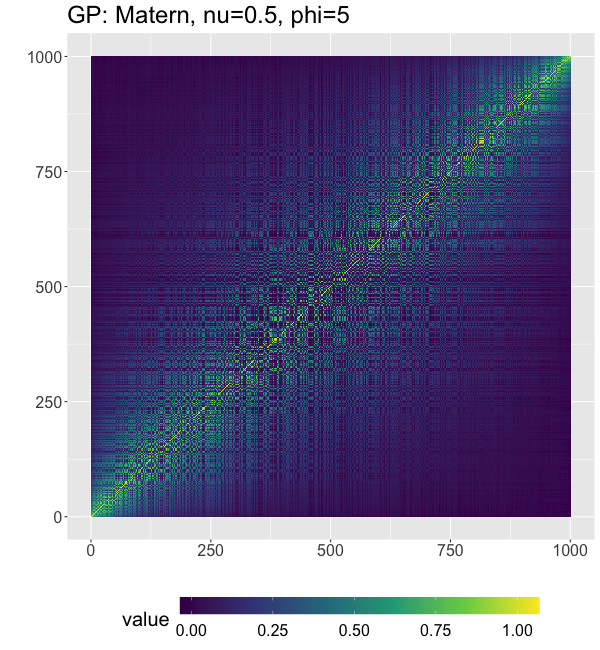}}
		\subfigure[NNGP: Exponential]{\includegraphics[scale=0.23,trim={0 0 0 30},clip]{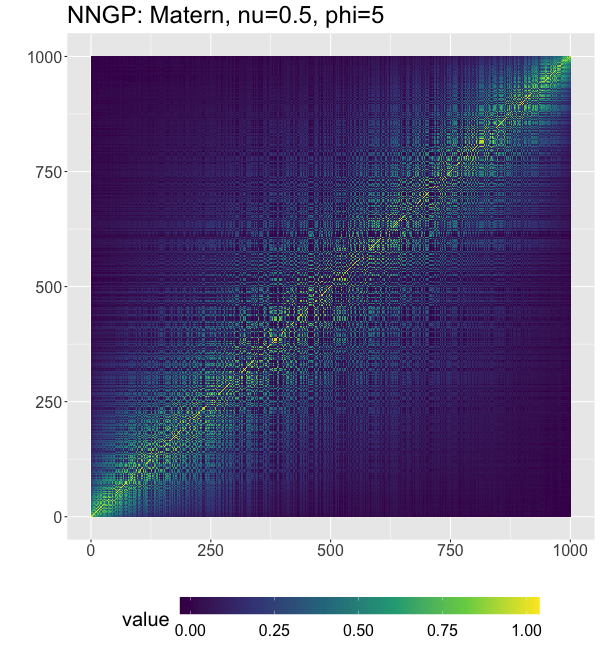}}
		\subfigure[Difference: Exponential]{\includegraphics[scale=0.23,trim={0 0 0 30},clip]{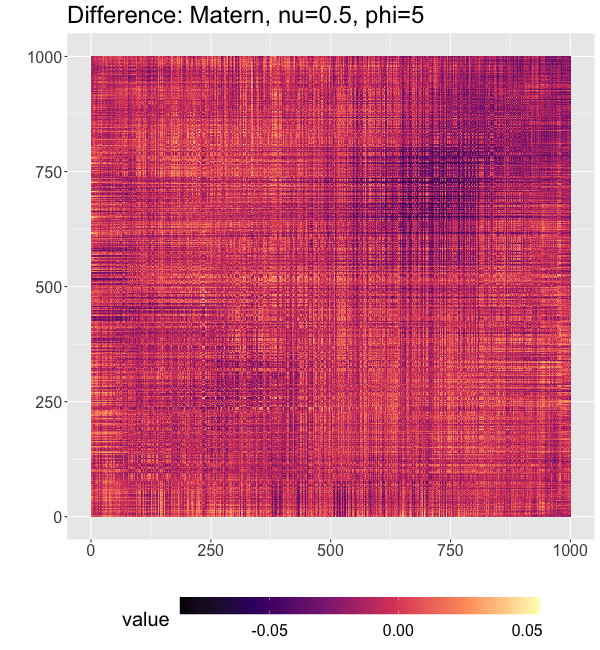}}
		\subfigure[GP: Mat\'ern $3/2$]{\includegraphics[scale=0.23,trim={0 0 0 30},clip]{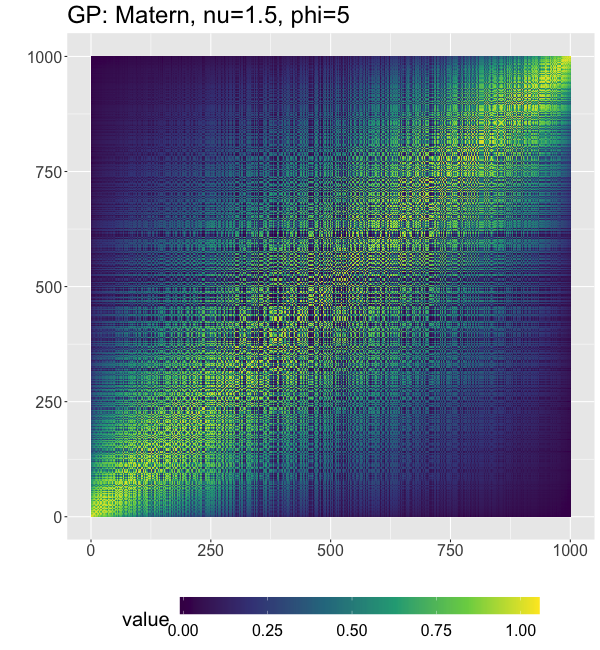}}
		\subfigure[NNGP: Mat\'ern $3/2$]{\includegraphics[scale=0.23,trim={0 0 0 30},clip]{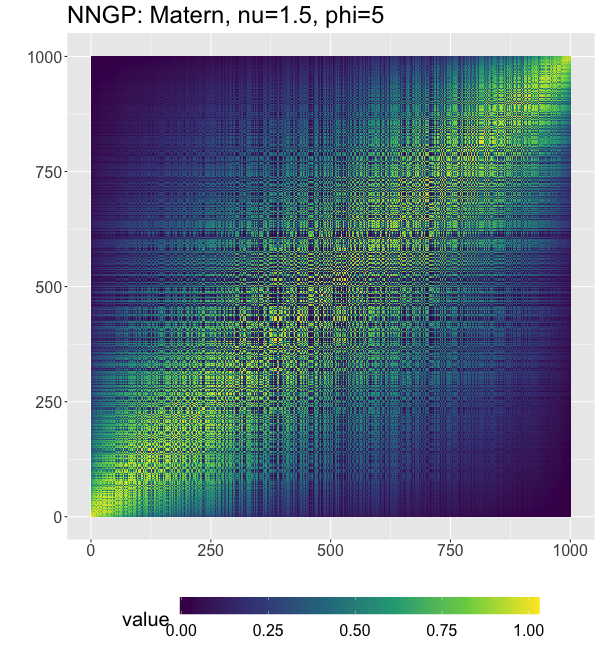}}
		\subfigure[Difference: Mat\'ern $3/2$]{\includegraphics[scale=0.23,trim={0 0 0 30},clip]{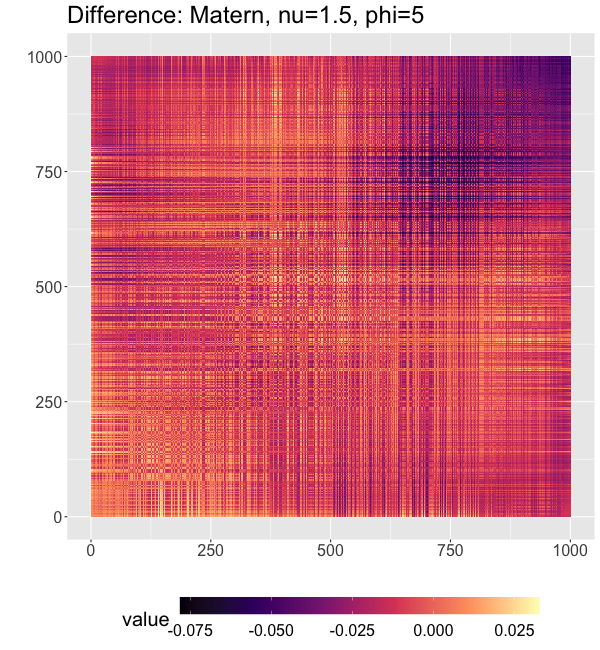}}
		\vskip-10mm \subfigure[Density of differences in sample covariances between full GP and NNGP]{\includegraphics[scale=0.75]{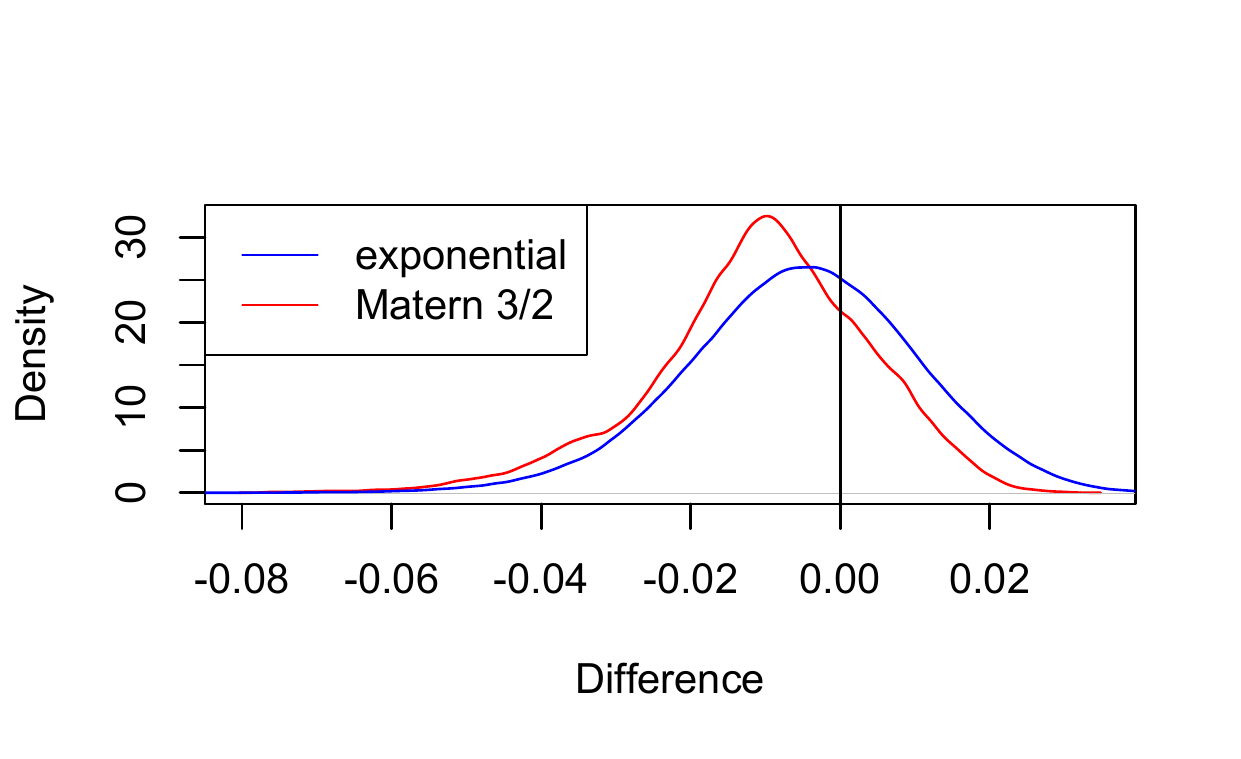}}
		\caption{\blue{Sample covariances of simulated} Gaussian random fields using (left column) full GP and (middle column) NNGP for the exponential covariance function (top row) and the Mat\'ern$_{3/2}$ covariance function (middle row). \blue{The right column plots the difference matrices between the two sample covariances.} The bottom row plots the density of the entries of these difference matrices between the sample covariance matrices generated from \blue{NNGP and full GP}.}
		\label{fig:sim}
	\end{figure}

	Figure \ref{fig:sim} assesses the quality of simulation of random fields using the full GP and NNGP for $n=1000$ randomly chosen locations and two choices of covariance functions (exponential and Mat\'ern$_{3/2}$). For each setting, $10000$ random draws are generated and the sample covariance matrix is plotted both for full GP and NNGP. We see that for each choice of the covariance function, the heat-maps of the sample covariance matrices look similar for the full GP and NNGP draws. Both methods display sharper decay for the exponential covariance function, and slower decay for the smoother Mat\'ern$_{3/2}$ function. For a more quantitative assessment of the approximation, we took the difference between the NNGP and full GP sample covariance matrices and plotted these matrices in the right column of Figure \ref{fig:sim} and the density of these differences in bottom row of Figure \ref{fig:sim}. We see that for both choices of covariance function, these densities peak around $0$ demonstrating the closeness of the approximation. \blue{However, we do see a pattern of small but consistent underestimation of the variances and covariances by NNGP. This issue is discussed more in Section \ref{sec:order}.}
	
	
	We also compare the run times for generating draws using full GP and NNGP in Table \ref{tab:sim}. We use the same two choices of covariance functions, sample sizes of $n=1000, 2500, 5000, 10000, 100000$. 
	\blue{For each sample size we generate $100$ draws, and repeat this experiment $20$ times per setting to report the means and standard deviations of run times across these $20$ runs. The personal computer used for this experiment had a 2.7 GHz Quad-Core Intel Core i7 processor and 16 GB of RAM.} We see that the NNGP draws take much less time than the corresponding full GP draws, and can be used for even the two largest sample sizes where we could not generate draws for the full GP as the process exhausts the computer memory, crashing the software. The cubic growth in the computing times for the full GP is clearly evident in the Table. For the NNGP, there is somewhat non-linear increase in the computing times as $n$ increases.  This is because, while the computation and storage times for NNGP are linear in $n$, it requires the neighbor sets as  inputs. There is a one-time cost of finding these sets of $m$ nearest neighbors for each location. This step can be time-consuming for large $n$, as detailed in \cite{finley2020r}. The times for NNGP presented in Table \ref{tab:sim} includes the timing for this neighbor-search step which explains this discrepancy. However, even with this expensive one-time operation, the simulation times for NNGP are impressive being able to generate \blue{$100$} draws of $100000$ realizations of the random field in about {2-3 minutes} minutes on a personal computer. 
	
	\begin{table}[h]
		\centering
		\begin{tabular}{cc|cc}
			Covariance function & Sample size    & NNGP & full GP \\ \hline
			\multirow{ 5}{*}{exponential} &  1000 &  0.7 (0.04)  &  2.6 (0.08) \\
			&  2500 &  1.6 (0.29)   &  31.8 (2.02) \\
			&  5000 &  3.3 (0.25)   & 262.3 (9.33) \\
			&  10000 &  8.3 (0.23)   &  NA \\
			&  100000 &  121.5 (9.53)   & NA \\ \hline
			\multirow{ 5}{*}{Mat\'ern$_{3/2}$} &  1000 &   1.6 (0.02)  &  2.0 (0.03) \\
			&  2500 &  4.3 (0.22)   & 34.6 (1.29) \\
			&  5000 &  8.2 (0.28)   & 266.6 (7.79) \\
			&  10000 &  16.7 (0.64)   & NA \\
			&  100000 &  202.8 (2.34)   & NA \\
		\end{tabular}
		\caption{{Means and standard deviations (within braces) of} computation times in seconds \blue{over 20 replicate experiments} for simulating {$100$} random draws from a full GP and NNGP.}
		\label{tab:sim}
	\end{table}

	
	Simulations from NNGP can now be performed using the \texttt{BRISC\_simulation} 
	function of the \texttt{BRISC} CRAN R-package \citep{briscpkg}.
	
	
	\subsection{Random forests with Gaussian Processes}\label{sec:rf}
	The hierarchical model (\ref{eq:gphier}) is a linear mixed-effects model with a linear covariate effect $\bx(\bs)'\bbeta$ and a spatial random effect $w(\bs)$ modeled as a GP. The effect size parameter $\bbeta$ helps infer about the association between the response and the covariates. A spatially varying coefficient (SVC) model \cite{gelfand2003spatial} allows the regression coefficient $\bbeta(\bs)$ to vary over space, but the hypothesized relationship between the response and the covariate in SVC model remains linear. 
	
	In recent years, owing to the progress of technology behind Geographical Information Systems (GIS), we have witnessed a deluge of large geospatial datasets in many research fields. These large datasets have offered the opportunity to relax the strong, \blue{and often inappropriate,} linearity assumption about the covariate effect. More general non-linear classes of functions $h(\bx(\bs))$ \blue{can model more complex relationships between the outcome and the covariates. The data-abundance in modern geo-spatial applications, allows} use of data-intensive machine learning methods for  estimation of such non-linear covariate effects. 
	A spatial non-linear mixed effects model will be given by 
	\begin{equation}\label{eq:gphiernl}
	\begin{aligned}
	y(\bs_i) = h(\bx(\bs_i)) + w(\bs_i) + \eps(\bs_i), \,
	\bw \sim N(0,\bC), \, 
	\eps(\bs_i) \iid N(0,\taus).  
	\end{aligned}
	\end{equation}
	
	Akin to (\ref{eq:gpmodel}), integrating out the spatial random effects leads to the marginal model 
	\blue{\begin{equation}\label{eq:gpmargnl}
	\begin{aligned}
	y(\bs_i) = h(\bx(\bs_i)) + \eps^*(\bs_i), \,  
	\end{aligned}
	\end{equation}
	where $\eps^*(\bs)$ is a dependent stochastic process with covariance function $\Sigma(\bs_i,\bs_j) = C(\bs_i,\bs_j) + \taus I(\bs_i = \bs_j)$.}
	
	Basis functions offer a natural avenue for such non-linear spatial mixed models. We can simply replace the linear covariate effect $\bx'\bbeta$ with $h(\bx)=B(\bx)'\bgamma$ where $B(\bx)$ is the basis expansion at $\bx$ for a choice of basis $B$ and $\bgamma$ are the unknown basis coefficients. As the problem remains linear in the parameters $\bgamma$, this generalization does not necessitate any change in estimation strategies. However, basis functions are generally not ideal to model certain classes of functions (e.g., functions with discontinuity like step-functions or piecewise continuous functions). More importantly, they suffer from the curse of dimensionality and are often inadequate if the number of covariates are greater than $3$ or $4$ \citep{taylor2013challenging}. 
	
	Random forests (RF) \citep{Breiman2001} has become one of the mainstays on non-parametric function estimation for \blue{non-linear and possibly non-smooth regression functions of the form (\ref{eq:gpmargnl}).} 
	RF has been shown to be a asymptotically consistent function estimation method \citep{scornet2015consistency} but the guarantees are established only under \blue{the assumption of iid errors $\eps^*_i := \eps^*(\bs_i)$. When (\ref{eq:gpmargnl}) arises from marginalization of a GP-distributed spatial random effect, the error process $\eps^*(\bs)$ will be dependent, and until recently, there were very few systematic studies on how data dependence affects performance of RF.}
	
	Despite this, RF has been used extensively in spatial or temporal settings where the data are correlated. Such applications have often ignored the data dependence, not using the spatial locations or the spatial covariance $\bSigma$, 
	and simply fitting the classic RF algorithm using only $\by$ and $\bX$. \cite{saha2020random} demonstrated that ignoring this spatial information significantly degrades the performance of RF. 
	\blue{Alternatively, some attempts to modify RF for spatial data has mostly abandoned the traditional GP-based mixed model framework of (\ref{eq:gphiernl}) or (\ref{eq:gpmargnl}) and have used additional distance/location based variables as extra covariates in the random forest fit. For example,} \cite{hengl2018random} developed a `{\em spatial random forests (spRF)}' that uses the paired distances between a location and all other locations as additional covariates. For $n$ locations, this adds $n$ extra covariates in the RF,  thereby leading to unnecessary dimension inflation \blue{ and  masking the effect of the $d$ true covariates when $ d\ll n$ \citep{saha2020random}. More importantly, these approaches, leave the additive setting of (\ref{eq:gphiernl}) and only model a joint covariate-spatial effect $E(y(\bs)) = g(\bx,\bs)$ and are thus suitable for predictions at new locations but cannot directly extract just the relationship between the outcome and the covariates.}
	
	\blue{There are many advantages of the GP-based linear (\ref{eq:gphier}) or non-linear (\ref{eq:gphiernl}) mixed-model frameworks. A complicated spatial effect $w(\bs)$ can be modeled flexibly using a GP specified by only 2 or 3 covariance parameters. The additive setting allows both estimating the covariate effect, and spatial predictions at new locations using kriging. Hence, for non-linear covariate effects, it is desirable create an RF algorithm that operates within this framework to account for the data dependence. However, unlike basis function methods which essentially reduces to a linear model in the basis function coefficient parameters facilitating direct optimization of a quadratic form log-likelihood, RF is a greedy localized algorithm and incorporating dependence across all data points within the RF algorithm is challenging.} 
	
	\cite{saha2020random} proposed estimating $h$ in the GP-based spatial non-linear mixed model (\ref{eq:gphiernl}) using a novel Random Forest algorithm that accounts for the spatial dependence. RF estimate is the average of many regression trees \citep{breiman1984classification}. The key observation to the extension of random forests to spatial data is that creation of a regression tree is equivalently characterized as a greedy ordinary least squares (OLS) optimization algorithm. To elaborate\blue{, consider the data model from (\ref{eq:gpmargnl}),  given by 
		\begin{equation}\label{eq:gpmodelnl}
		\by \sim N(\bh,\bSigma)  \mbox{ where } \bh=(h(\bx(\bs_1)),\ldots,h(\bx(\bs_n)))'.
		\end{equation}}
	A tree estimate (with $K$ leaf nodes) of $\bh$ can be represented using a $n \times K$ binary membership matrix $\bZ$ whose $(i,j)^{th}$ element is $1$ if the $i^{th}$ observation belongs to the $j^{th}$ leaf-node, and a vector $\bbeta=(\beta_1,\ldots,\beta_K)'$ where $\beta_k$ is the value of the estimate representing the $k^{th}$ node. 
	Thus at the $r^{th}$ iteration of the algorithm, $\bh$ is modeled as  $\bZ^{(r)}\bbeta^{(r)}$ for this membership matrix $\bZ^{(r)}$  and the representative values $\bbeta^{(r)}$. To find the best choices of $\bZ^{(r)}$ and $\bbeta^{(r)}$, one optimizes the OLS loss: 
	\begin{equation}\label{eq:rfols}
	(\widehat{\bZ^{(r)}},\widehat{\bbeta^{(r)}}) = \underset{\bZ^{(r)} \in \mathcal C^{(r)}, \bbeta^{(r)}}{\arg \min} \|\bY - \bZ^{(r)} \bbeta^{(r)}\|_2^2.
	\end{equation}
	Here $\mathcal C^{(r)}$ is the class of eligible design matrices given the previous set of nodes in the tree and the covariate values. The optimal $\widehat{\bZ^{(r)}}$ gives the next set of nodes created in the tree, while the corresponding $\widehat{\bbeta^{(r)}}$ gives the updated set of node representative values. This process of tree creation is continued till a termination criterion is met (maximum number of nodes, minimum number of members per node, etc.). 
	
	\cite{saha2020random} noted that the OLS optimization in (\ref{eq:rfols}) is equivalent to obtaining MLE for $(\bZ^{(r)},\bbeta^{(r)})$ from the model $\by \sim N(\bZ^{(r)} \bbeta^{(r)},\bI)$. Under spatial dependence, we have Cov$(\by)=\bSigma$. To incorporate this spatial covariance, one should consider the model (\ref{eq:gpmodelnl}) with $\bh=\bZ^{(r)} \bbeta^{(r)}$, i.e., 
	$\by \sim N(\bZ^{(r)} \bbeta^{(r)},\bSigma)$. Thus, for a given $\bSigma$, they proposed optimizing a generalized least square (GLS) loss 
	\begin{equation}\label{eq:rfgls}
	(\widehat{\bZ^{(r)}},\widehat{\bbeta^{(r)}}) = \underset{\bZ^{(r)} \in \mathcal C^{(r)}, \bbeta^{(r)}}{\arg \min} (\bY - \bZ^{(r)} \bbeta^{(r)})'\bSigma^{-1}(\bY - \bZ^{(r)} \bbeta^{(r)}).
	\end{equation}
	
	The new GLS loss accounts for the spatial dependence in the data. For the optimal choice of $\widehat{\bZ^{(r)}}$, the node representatives will be given by the GLS estimate
	\begin{equation}\label{eq:rfglsbeta}
	\widehat{\bbeta^{(r)}} = \left(\widehat{\bZ^{(r')}}\bSigma^{-1}\widehat{\bZ^{(r)}}\right)^{-1}\widehat{\bZ^{(r')}}\bSigma^{-1}\by.
	\end{equation}
	
	The optimization criterion (\ref{eq:rfgls}) and the node representatives (\ref{eq:rfglsbeta}) complete the algorithm for one regression tree. To create a forest from trees, in iid settings, data are resampled and a regression tree is estimated for each resampled dataset, which are then averaged to create the forest estimate. However, as discussed in Section \ref{sec:boot}, direct resampling of spatially correlated data violates assumptions of bootstrap, and it is unclear what $\bSigma$ would be for a resampled dataset as the correlation between two resamples of the same data unit $(y_i,\bx(\bs_i),\bs_i)$ would be $1$. 
	
	The GLS approach to random forests offers a synergistic solution to the resampling problem. A GLS regression between $\by$ and some design matrix $\bZ$, with a covariance matrix $\bSigma$, can be thought of as an OLS regression between the decorrelated vector $\by^*=\bSigma^{-1/2}\by$ and $\bSigma^{-1/2}\bZ$. As in Section \ref{sec:boot}, for spatial data, it makes sense to resample the decorrelated vector $\by^*$. Thus, for a resampling matrix $\bP_t$ used to create the $t^{th}$ regression tree, we can simply modify (\ref{eq:rfgls}) for the resample and use 
	\begin{equation}\label{eq:rfglsre}
	(\widehat{\bZ^{(r)}},\widehat{\bbeta^{(r)}}) = \underset{\bZ^{(r)} \in \mathcal C^{(r)}, \bbeta^{(r)}}{\arg \min} (\bY - \bZ^{(r)} \bbeta^{(r)})'\bSigma^{-T/2}\bP_t'\bP_t\bSigma^{-1/2}(\bY - \bZ^{(r)} \bbeta^{(r)}).
	\end{equation}
	for partitioning of the nodes. The corresponding node representatives will now be given by 
	\begin{equation}\label{eq:rfglsbetare}
	\widehat{\bbeta^{(r)}} = \left(\widehat{\bZ^{(r)'}}\bSigma^{-T/2}\bP_t'\bP_t\bSigma^{-1/2}\widehat{\bZ^{(r)}}\right)^{-1}\widehat{\bZ^{(r)'}}\bSigma^{-T/2}\bP_t'\bP_t\bSigma^{-1/2}\by.
	\end{equation}
	
	\cite{saha2020random} referred to this new random forest algorithm as `{\em RandomForestsGLS or RF-GLS}' -- owing to its strong connection to generalized least squares. Using numerical studies across a wide span of settings, they showed that RF-GLS substantially improves over naive RF for estimating the mean function $h$, and over both naive RF and spRF of \cite{hengl2018random} for spatial prediction.
	
	\blue{Sparse Cholesky factors play a central role in ensuring an efficient implementation of RF-GLS.} Note that both the partitioning criterion (\ref{eq:rfglsre}) and the node values (\ref{eq:rfglsbetare}) requires computing the Cholesky factor $\bSigma^{-1/2}$. \blue{While this is a one-time computation, this would require $O(n^3)$ FLOPs and is thus not feasible for large $n$. Additionally, both (\ref{eq:rfglsre}) and (\ref{eq:rfglsbetare}) involve multiple evaluations of quadratic forms of $\bSigma^{-T/2}\bP_t'\bP_t\bSigma^{-1/2}$. Since the optimal partition among the candidate ones is found in a brute force manner in regression trees, these quadratic forms are reevaluated for all possible node-partitions. Thus even with a pre-computed Cholesky factor $\bSigma^{-1/2}$, such repeated dense matrix operations are impossible.} Hence, for practical implementation once again the NNGP Cholesky factor $\tSig^{-1/2}$ is used in all steps thereby ensuring linear time and storage for the algorithm. 
	
	\blue{\cite{saha2020random} also developed a comprehensive theory of asymptotic $L_2$-consistency of RF-GLS in estimating the non-linear function $h$ in (\ref{eq:gphiernl}). To our knowledge, this was the first theory for forest and tree estimators under dependence. Consistency of GLS-style regression trees and RF-GLS was established for a wide-range of $\beta$-mixing dependent error processes. This class of processes include the popular Mat\'ern GP for spatial data and autoregressive time-series. }
		
	\blue{Sparse Cholesky factors also play a key-role in the theoretical results. Consistency of RF-GLS currently requires a regular sparse Cholesky factor of the working precision matrix $\Sigma^{-1}$ even if the data is generated from a process yielding a dense covariance matrix. For equi-spaced 1-dimensional spatial locations NNGP Cholesky factors satisfy this regularity condition and ensures consistency of RF-GLS even when the true data is generated from a full Mat\'ern GP. Similarly, Cholesky factors of auto-regressive covariances are sparse, leading to the consistency result for such temporal processes.}
	
	 The method 
	 has been implemented in the R-package \texttt{RandomForestsGLS} \citep{rfgls}. 
	
	\subsection{Directed acyclic graph autoregression (DAGAR) for areal data}\label{sec:areal}
	
	Most applications of sparse Cholesky matrices in spatial statistics, as discussed in this review, pertain to the paradigm of point-referenced or geospatial data. However, the idea has also been exploited to recently propose a new class of generative and interpretable models for areal (geographically-aggregated) data. Each unit in such a dataset represents a geographical region (e.g., counties or states). The data can be envisioned as consisting of triplets $(y_i,X_i,R_i)$ where $R_i$ is the $i^{th}$ region, and $y_i$ and $\bx_i$ are the corresponding response and covariates. A generalized mixed-linear model commonly used to analyze such data is specified as:
	\begin{equation}\label{eq:areal}
	g(E(y_i)) = \bx_i'\bbeta + w_i, i=1,\ldots,n
	\end{equation}
	where $g$ is a suitable link function for the outcome type, $\bx_i'\bbeta$ is the fixed covariate effect and $w_i$ is the area-specific random effect. The random effect vector $\bw=(w_1,\ldots,w_n)'$ is typically endowed with multivariate Gaussian prior with the covariance (or precision) matrix capturing the geographical information. 
	
	Unlike point-referenced data where location of each unit is a $2$- or $3$-dimensional co-ordinate vector, it is challenging to encapsulate information of an entire geographical area $R_i$ by some representative co-ordinates. Hence, a typical approach to capturing spatial structure in areal data is to represent the geography in terms of a graph $\mathcal G$ with vertices as the set of regions, and the adjacency matrix $\bA=(A_{ij})$ specified to capture the relative spatial information in the regions. One common way of doing this is setting $A_{ij}=1$ if regions $i$ and $j$ share a border. Subsequently, to impose smoothness across neighboring regions (which share an edge in the graph), a popular prior for $\bw$ is the `{\em Conditional Autoregressive (CAR)}' model
	\begin{equation}\label{eq:car}
	\bw \sim N(0, \sigs (\bD - \rho \bA)^{-1})
	\end{equation}
	where $\bD=$diag$(m_1,\ldots,m_n)$ is a diagonal matrix with $m_i$ being the number of neighbors of $R_i$, $\sigs$ is the marginal variance parameter, and $\rho$ controls the degree of spatial smoothness. 
	
	The CAR model was originally proposed by \cite{besag74} via the conditional distributions
	\begin{equation}\label{eq:carcond}
	w_i \given \bw_{-i} \sim N(\frac 1{m_i} \sum_{j \sim i} w_j, \frac {\sigs}{m_i})
	\end{equation}
	where $\bw_{-i}=(w_1,\ldots,w_{i-1},w_{i+1},\ldots,w_n)'$ and $j \sim i \iff A_{ij}=0$.  
	The conditional model can be shown to be equivalent to (\ref{eq:car}) for $\rho=1$ and offers a much clear insight into why this model effectuates spatial smoothing. The conditional mean of $w_i$ is the mean of its neighbors, and the conditional precision to be proportional to the number of neighbors. However, for $\rho=1$, the precision matrix $\bD - \bA$ is singular for all graphs. This is an improper prior and is referred to as the `{\em Improper or Intrinsic CAR (ICAR)}'. The ICAR model was later extended to include the parameter $\rho < 1$ in (\ref{eq:car}) such that the resulting precision matrix is positive definite. This creates a proper prior and can model a wider range of spatial structures. 
	
\begin{figure}[]
		\centering
		\includegraphics[scale=0.4]{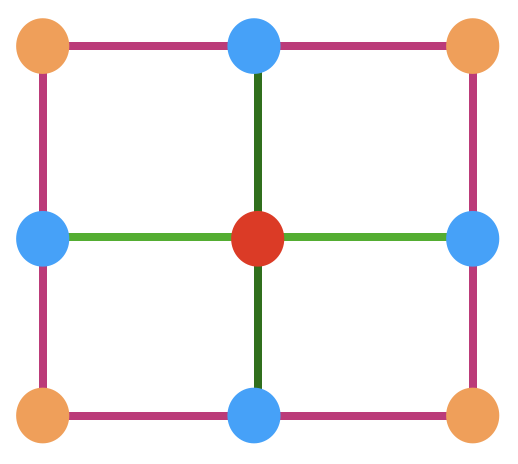}
		\includegraphics[scale=0.15]{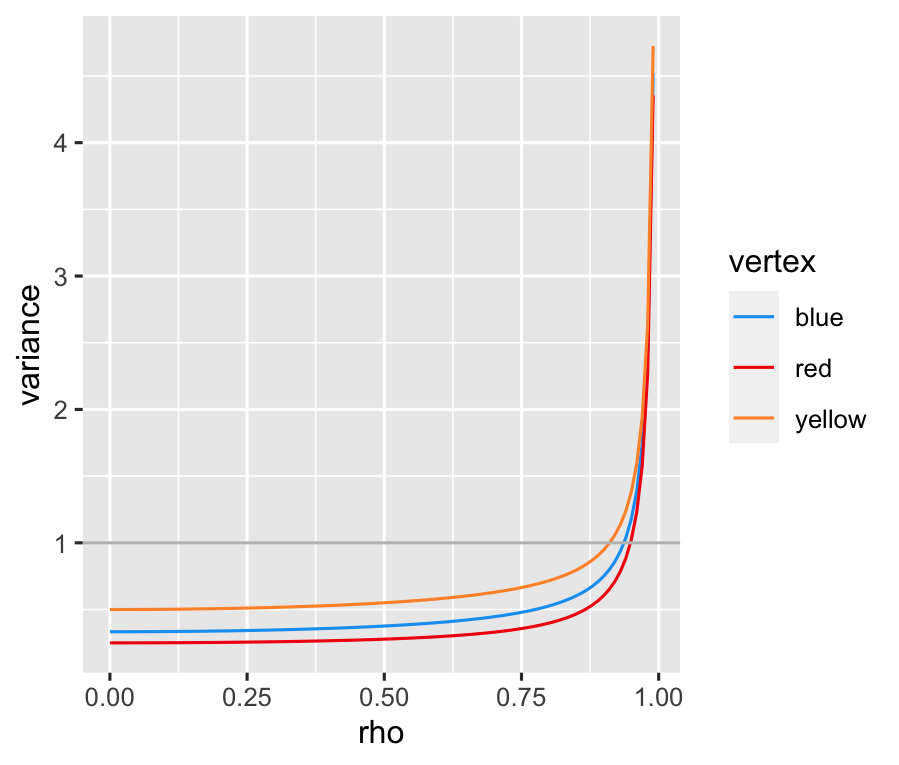}
		\includegraphics[scale=0.15]{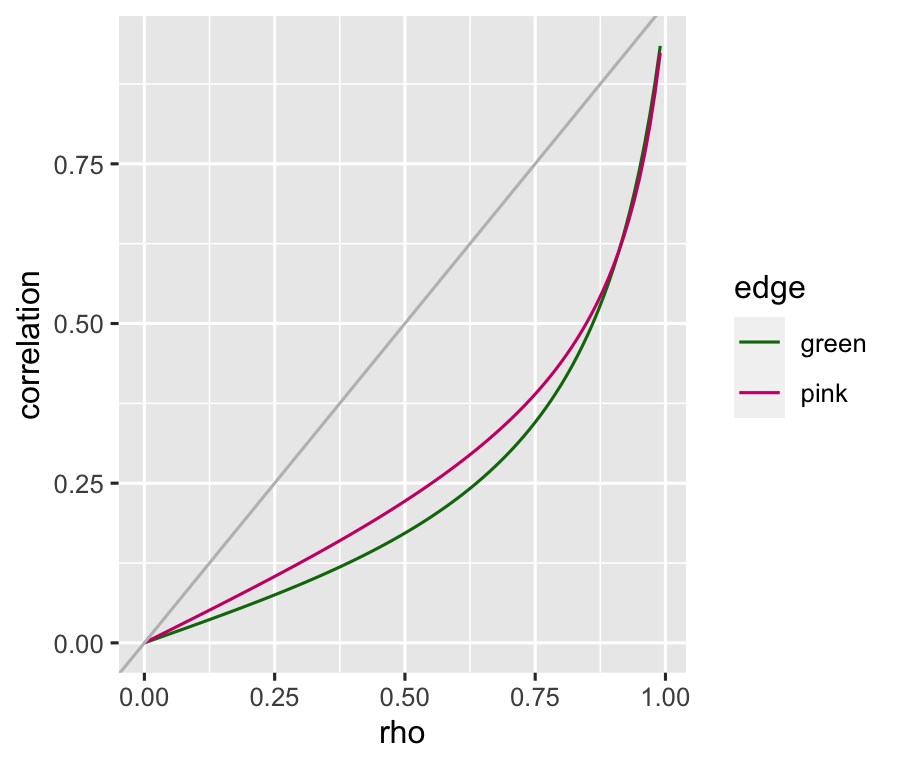}
		\caption{Unequal variances and correlations in the CAR model:  (left) A $3 \times 3$ grid graph with vertices and edges grouped and colored by symmetry, (middle) CAR-induced variances for each vertex group as a function of $\rho$, (right) CAR-induced neighbor-pair correlations for each edge group as a function of $\rho$.} 
		\label{fig:car}
	\end{figure}

	The CAR model has drawn criticisms regarding interpretability of the parameter $\rho$ and its complicated association with the spatial correlation induced \citep{wall04,assun09,banerjee2014hierarchical}. In addition, we also note that the resulting covariance matrix from the CAR model is heteroskedastic. To illustrate these issues, we study the CAR covariance matrix $(\bD - \rho \bA)^{-1}$ for a simple $3\times 3$ grid graph plotted in Figure \ref{fig:car} (left). The vertices and edges are assigned the same color group if their relative orientations are symmetric (e.g., all corner vertices are symmetric to each other). Figure \ref{fig:car} (middle) then plots the CAR variance for each vertex group as a function of $\rho$. We see that the variances are not same across groups, i.e, CAR model is not homoskedastic. Regions with higher number of neighbors are endowed with lower marginal variance. While it is reasonable to model the conditional variance of a unit (given its neighbors) to be less if there are more neighbors, the same logic should not be true of the marginal variance. Hence, this heteroskedasticity is problematic. Figure \ref{fig:car} (right) plots the correlation between neighboring units as a function of $\rho$. We see that the correlation is again not same for each neighbor-pair and that the relationship between this correlation and $\rho$ is highly non-linear. This prohibits any direct interpretation of $\rho$ as has been highlighted in the aforementioned studies. 
	
	\cite{dagar} proposed a new class of models based on sparse Cholesky factors that circumvents these issues of CAR models. The construction is identical to  (\ref{eq:nngpgen}) and (\ref{eq:chol}) which can be viewed as a general method to create multivariate Gaussian distributions via Cholesky factors. The specific choices of $\bB$ and $\bF$ lead to different models. In the context of geo-spatial modeling discussed in the previous sections, the central step was to select the neighbor sets. Given the neighbor sets, the matrices $\bB$ and $\bF$ are automatically determined from the covariance matrix $\bC$ of the full GP joint distribution which the NNGP is approximating. 
	
	For areal data, the neighbor sets are naturally defined by the graph $\mathcal G$ as regions sharing an edge in the adjacency matrix. Hence, \cite{dagar} proposed setting $b_{ij}=0$ in (\ref{eq:nngpgen}) unless $i \sim j$ and $j < i$. However, the central choice here involves specifying the non-zero $b_{ij}$'s and $f_i$'s. Unlike NNGP which used the covariance matrix $\bC$ from the full GP covariance, for areal data there is no suitable joint distribution from which to derive $\bB$ and $\bF$. Using the CAR covariance matrix would lead to an approximation of it that inherits its aforementioned drawbacks. If the graph was a tree (acyclic graph), then a valid candidate to derive $\bB$ and $\bF$ is the autoregressive covariance on the tree, i.e., $\bC=(\rho^{d_{ij}})$, where $d_{ij}$ denotes the shortest path between vertices $i$ and $j$ on the graph \citep{mar}. However, graphs generated by specifying edges between geographical regions sharing a border will typically not be acyclic and such a $\bSigma$ will not be positive definite. One option would be to approximate the whole graph with a minimal spanning tree (MST) but such an approximation leads to large errors by leaving out many edges \citep{suddtreethesis}.
	
	\cite{dagar} noted that specifying the $i^{th}$ row of $\bB$ and $f_i$ only requires a working assumption on the joint distribution of the $w_k$'s on the sub-graph $\mathcal G_i$ corresponding to the vertices $i \cup \{j \given j \sim i, j < i\}$. They then used an autoregressive covariance matrix on the MST $\mathcal T_i$ of this sub-graph. $\mathcal T_i$ has only edges between $i$ and all $j \sim i, j < i$. This leads to thee following choices:
	\begin{equation}\label{eq:bf}
	b_{ij} = I(j \sim i, j < i) \frac \rho{1+(m_{<i} - 1)\rho^2}, f_i = \frac{\sigs}{1+(m_{<i} - 1)\rho^2},
	\end{equation}
	where $m_{<i} = |\{j \given j \sim i, j < i\}|$ is the number of directed neighbors of $i$. There are several advantages of this choice. Unlike approximating the entire graph $\mathcal G$ with an MST, sequentially approximating each subgraph $\mathcal G_i$ 
	with $\mathcal T_i$ ensures that no edge in the original graph is disregarded as any edge $i \sim j$ will be in either $\mathcal T_i$ or $\mathcal T_j$. Use of the autoregressive working covariance is justified by its desirable property of having homoskedastic variances and same neighbor-pair correlation $\rho$. We will note subsequently that the resulting areal model inherits these properties. Also, note that in the limit of $\rho \to 1$, the $i^{th}$ conditional distribution resulting from (\ref{eq:bf}) becomes
	$$ w_i \given w_1, \ldots, w_{i-1} \sim N(\frac 1{m_{<i}}\sum_{j < i, j \sim i} w_j, \frac \sigs {m_{<i}}).$$
	This is similar to the conditional distributions (\ref{eq:carcond}) in the ICAR model but using the directed neighbors instead of all neighbors. 
	
	\cite{dagar} referred to the model as `{\em Directed acyclic graph autoregression (DAGAR)}' model because of its reliance on autoregressive covariances and as the undirected graph $\mathcal G$ combined with the ordering imposed to construct the sparse Cholesky factor essentially yields a directed acyclic graph. Several nice properties of DAGAR model has been established with regards to parameter interpretability and computation. For any tree or grid graph, DAGAR covariance matrices was shown to be homoskedastic, and $\rho$ could be directly interpreted as the exact correlation between each pair of neighboring regions. If the graph $\mathcal G$ has $e$ edges, the storage and computational cost of evaluating the DAGAR likelihood is $O(n+e)$, same as the sparsity of the Cholesky factor $\bF^{-1/2}(\bI - \bB)$ or equivalently of $\bB$. Simulation studies covering a wide range of scenarios showed that DAGAR models tend to perform better than CAR models for low or moderate spatial correlation in the data, while the performance of the two classes of models are similar in presence of strong spatial correlation. The model has subsequently been generalized to include multivariate outcomes \citep{gao2019spatial,gao2021hierarchical}. 
	


	\section{Discussion} \label{sec:disc}
	Vecchia's approximation, NNGP and the resulting sparse Cholesky matrices have become increasingly popular in recent years owing to their spectacular empirical success in mimicking inference from full Gaussian Processes while being extremely scalable. In this manuscript we reviewed these methods and discussed in greater details some recent and diverse applications of these sparse Cholesky based beyond their tradition usage in scalable estimation and prediction for geo-spatial data. We conclude the manuscript with discussion of some yet-to-be-addressed aspects  of this approach which opens up potential avenues of future research. 
	
	\subsection{Loss of homoskedasticity}\label{sec:order} 
	As discussed in Section \ref{sec:litrev}, the sparse Cholesky factors generated from NNGP rely on a predetermined ordering of the locations. The order dependence is evident from the construction in (\ref{eq:chol}) and can affect the performance of the method in terms of parameter efficiency \citep{guinness2018permutation}. 
	For a stationary covariance function $C$, the full GP covariance matrix $\bC$ is homoskedastic, i.e., $C_{ii}=\sigs$, the spatial variance. The impact of NNGP approximation on homoskedasticity has not been studied. 
	We here show empirically that the NNGP precision matrices are not homoskedastic in general and that the degree of departure from homoskedasticity can be order-dependent. 
	
\begin{figure}[]
		\centering
		\includegraphics[scale=0.2]{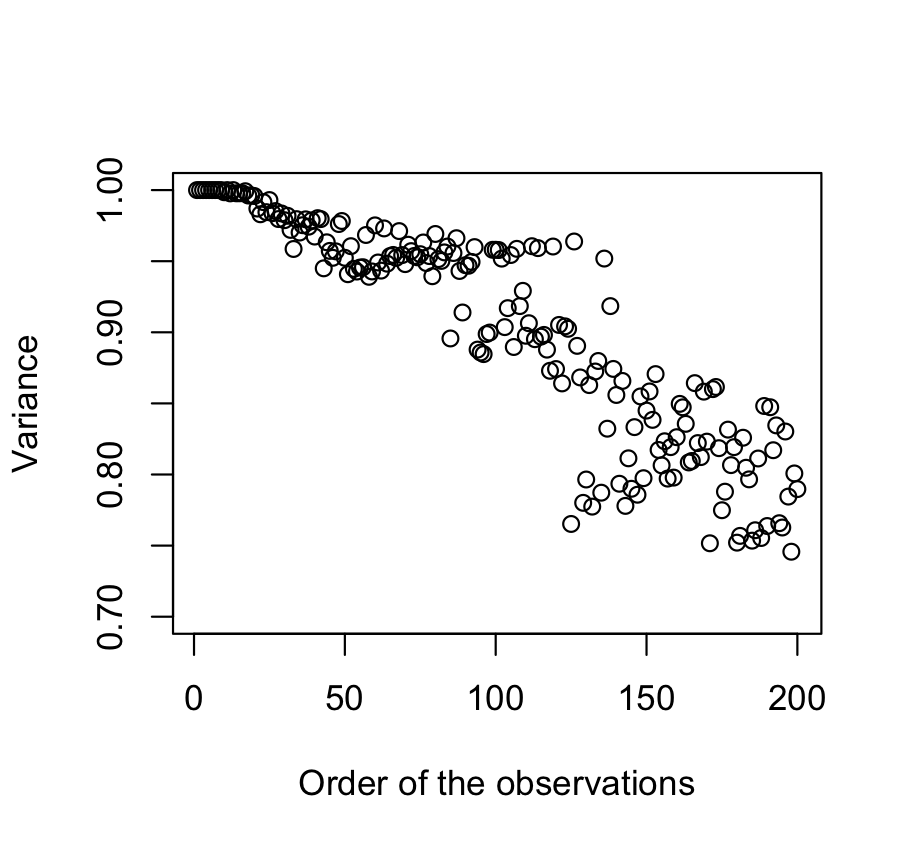}
		\includegraphics[scale=0.2]{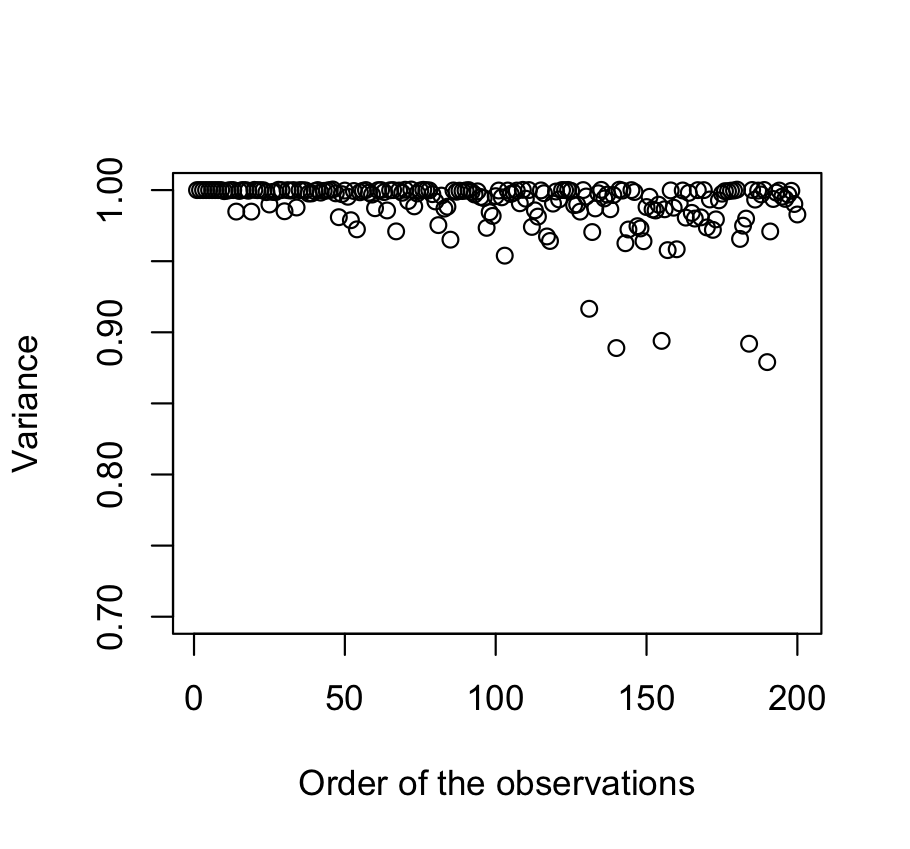}
		\caption{Spatial variances in NNGP plotted against order-index of the location for co-ordinate based ordering (left) and random ordering (right).} 
		\label{fig:hom}
	\end{figure}

	We generated $200$ locations randomly on an unit square, and use an exponential covariance function with $\sigs=1$ and $\phi=1$, and $m=5$ nearest neighbors. Figure \ref{fig:hom} plots the variance of $w_i$'s against the ordering $i$ where $\bw=(w_1,\ldots,w_n) \sim N(0,\tC)$ is simulated from an NNGP model. The left figure is for co-ordinate based ordering and we see that with increase in the location order $i$ there is a sharp decline in the NNGP variances from the true variance of $1$. For units ordered towards the very end, these variances are as low as $0.75$. The right figure is for random ordering. The lack of homoskedasticity is less prominent here with worst case variance of around $0.9$. However, both the magnitude and frequency of decline in the NNGP variances still generally increase with the order of the unit $i$ (i.e., towards the right-side of the $x$-axis). 
	
	These results show that NNGP variances often underestimate the full GP variance, and the extent of this is heavily dependent on the choice of ordering. This  under-estimation of the variance can potentially manifest itself in aggressive prediction intervals with poor coverage for units ordered towards the end. If the ordering is geographical (like co-ordinate based) this may imply that the quality of prediction can degrade in regions corresponding to higher orderings. For random ordering, as the units ordered towards the end will be spread out randomly throughout the whole region, quality of prediction intervals for a specific region is less likely to be impacted although overall prediction quality may still suffer. Finally, as the number of neighbors $m$ is increased, the quality of NNGP approximation improves and for all choices of ordering, the NNGP variances are much closer to $1$. 
	
	\blue{Alternate approaches for big spatial data like ones relying on latent realizations of the process on a set of knots (fixed locations like a grid), like the modified predictive process \citep{finley2009} or the multi-resolutional approximation \citep{katzfuss2017multi}, homoskedasticity is preserved. These methods belong to the general class of sparse generalized Vecchia approximations \citep{katzfuss2021}. However, for the original Vecchia's approximation \citep{ve88} or NNGP that are constructed without relying on any latent knots, it is still unknown if there exists an ordering and neighbor-selection strategy that can exactly preserve homoskedasticity.}
	
	\subsection{Neighbor selection for GP covariance functions on multi-dimensional non-spatial domains}\label{sec:sep} 
	Gaussian Processes are widely used as stochastic emulators of computer outputs \citep{currin1991bayesian} where the `space' is not physical space but some abstract multi-dimensional parameter domain. \blue{Similarly, GPs are popular as non-parametric function estimators on multi-dimensional covariate spaces.} In such cases the `{\em locations}' $\bs$ are high-dimensional  parameter or covariate values and each co-ordinate of the location (corresponding to a different parameter or covariate) has a difference scale of variation. The choice of Euclidean distance-based neighbor sets used in Vecchia's approximation and NNGP is rooted in the assumption of isotropic covariance functions. These covariance functions ignore the relative scales of each co-ordinate of the location vector and are not suitable \blue{for such tasks on non-spatial domains disparate scales along each dimension.}
	
	Separable covariance functions that endow each co-ordinate with its own scale parameter, are widely popular for such applications of GP using high-dimensional non-spatial inputs (locations). If  $\bs=(s^{(1)},\ldots,s^{(K)})'$ denote a $K$-dimensional location, then a separable covariance function between two such locations $\bs_i$ and $\bs_j$ is given by:
	\begin{equation}\label{eq:sep}
	C(\bs_i,\bs_j \given \btheta) = \prod_{k=1}^K C_k(s_i^{(k)},s_j^{(k)} \given \btheta_k),
	\end{equation}
	where $C_k$'s, \blue{parametrized by $\btheta_k$,} are covariance functions for one-dimensional \blue{domains, and $\btheta$ stacks up the $\btheta_k$'s for all the covariates.} For example, if each $C_k$ is chosen from the popular Gaussian family of covariances with decay parameter $\phi_k$, the resulting \blue{anisotropic} covariance function is given by:
	\begin{equation}\label{eq:sepgauss}
	C(\bs_i,\bs_j \given \btheta) = \sigs \exp\left( - \sum_{k=1}^K \phi_k (s_i^{(k)}-s_j^{(k)})^2 \right).
	\end{equation}
	
	\blue{A related class of kernels for GPs used in function emulation are the automatic relevance determination (ARD) kernel \citep{kailong2020} specified as 
	\begin{equation}\label{eq:ard}
	C(\bs_i,\bs_j \given \btheta) = C_{iso}( \| \widetilde \bs_i - \widetilde \bs_j \| \given \btheta_0)
	\end{equation}
	where $C_{iso}$ is an isotropic covariance kernel parametrized by $\btheta_0$, and $\btheta=(\btheta_0,\blambda)'$ where $\blambda=(\blambda_1,\ldots,\blambda_K)'$ is a vector of scale parameters such that for any location $\bs=(s^{(1)},\ldots,\bs^{(K)})'$, the scaled location $\widetilde \bs$ is defined as $\widetilde \bs=(s^{(1)}/\lambda_1,\ldots,\bs^{(K)}/\lambda_K)'$. If $C_{iso}$ is the isotropic Gaussian kernel, then the ARD kernel agrees with the separable kernel (\ref{eq:sepgauss}) with $\phi_k = 1/\lambda_k^2$. }
	
	It is unclear how to efficiently guide neighbor set selection of NNGP for such separable \blue{or ARD} covariance functions on high-dimensional input spaces. 
	Euclidean-distance based nearest neighbors of a given location no longer have the highest correlation with that location. The correlation contour is now highly dependent on the values of the decay parameters $\phi_1, \ldots, \phi_K$ \blue{($\lambda_k$'s for the ARD kernels)}. For example, if $\phi_1 \gg \phi_k$ for all $k \neq 1$, selecting neighbors based on the first co-ordinate than the total Euclidean distance will lead to a much better approximation. As these parameters $\phi_k$'s are unknown, it is challenging to set such a strategy apriori. 
	
	One can pursue a dynamic neighbor-selection strategy, searching for locations that have the highest correlation with a given location for each new value of the $\phi_k$'s. However, such brute-force neighbor searches at each iteration of an MCMC or optimization algorithm will drastically slow down the method. \cite{datta2016nonseparable} explored this problem in the context of spatio-temporal models, which also uses anisotropic covariance functions as variation along space and time also has different scales. They proposed a more restricted search perimeter that is guaranteed to contain the $m$ highest-correlated neighbors. Whether such a strategy computationally tenable for  high-dimensional input spaces needs to be researched. \blue{\cite{katzfuss2020scaled} used  nearest-neighbor based sparse Cholesky factors of ARD covariance matrices in GP-based function emulation. They updated the parameters $\btheta_0$ and $\blambda$ at every iteration, but to circumvent dynamic neighbor-selection at each iteration, they updated the ordering and neighbor sets only at certain iterations separated by exponential increasingly gaps.}
	As input points are further spread apart in higher dimensions, research also needs to be conducted to investigate how the established trade-offs between the quality of nearest neighbor (or highest-correlation) approximations and the size of the neighbor set hold up for such high-dimensional locations.  

\section*{Acknowledgements}
	The author thanks two anonymous reviewers for their suggestions that helped improve the manuscript, and Dr. Matthias Katzfuss for helpful discussions on related work. The author was supported by NSF award DMS-1915803.
	
	\bibliographystyle{asa}
	\bibliography{refs}
	
\end{document}